\documentclass[aps,prd,twocolumn,twoside,superscriptaddress,floatfix, nofootinbib,10pt]{revtex4-1}
\pdfoutput=1
\usepackage{graphicx}
\usepackage{color}
\usepackage[colorlinks,citecolor=blue,urlcolor=blue]{hyperref}
\usepackage{xspace}
\usepackage{nicefrac}
\usepackage{empheq}
\usepackage{booktabs}
\usepackage{float}
\AtBeginDocument{
\heavyrulewidth=.08em
\lightrulewidth=.05em
\cmidrulewidth=.03em
\belowrulesep=.65ex
\belowbottomsep=0pt
\aboverulesep=.4ex
\abovetopsep=0pt
\cmidrulesep=\doublerulesep
\cmidrulekern=.5em
\defaultaddspace=.5em
}

\newcommand{\beq} {\begin{equation}}
\newcommand{\eeq} {\end{equation}}
\newcommand{\bal} {\begin{aligned}}
\newcommand{\eal} {\end{aligned}}

\usepackage{aas_macros}
\usepackage[utf8]{inputenc}
\usepackage{ifthen,amsmath,amssymb, bm}

\newcommand{\Clpp}{\ensuremath{C_L^{\phi\phi}}\xspace}
\newcommand{\hClpp}{\ensuremath{\widehat C_L^{\phi\phi}}\xspace}
\newcommand{\op}[1]{\mathbb{#1}}
\newcommand{\Len}[1][]{\op L\ifthenelse{\equal{#1}{}}{}{(#1)}}
\newcommand{\Cflen}[1][]{\op{\widetilde{C}}_{f}\ifthenelse{\equal{#1}{}}{}{(#1)}}
\newcommand{\Cf}[1][]{\op C_{f}\ifthenelse{\equal{#1}{}}{}{(#1)}}
\newcommand{\Cphi}[1][]{\op C_{\phi}\ifthenelse{\equal{#1}{}}{}{(#1)}}
\newcommand{\Cn}{\op C_{n}}
\newcommand{\Cg}{\op C_{g}}
\newcommand{\Scut}{S_{\rm cut}}

\hypersetup{colorlinks,linkcolor={magenta},citecolor={magenta},urlcolor={magenta}}

\begin{document}

\title{
Impact \& Mitigation of Polarized Extragalactic Foregrounds on Bayesian Cosmic Microwave Background Lensing
}

\author{Frank J. Qu}\email{jq247@cantab.ac.uk}
\affiliation{DAMTP, Centre for Mathematical Sciences, University of Cambridge, Wilberforce Road, Cambridge CB3 OWA, UK}
\affiliation{Kavli Institute for Cosmology Cambridge, Madingley Road, Cambridge CB3 0HA, UK}
\affiliation{Kavli Institute for Particle Astrophysics and Cosmology, 382 Via Pueblo Mall Stanford, CA 94305-4060, USA}

\author{Marius Millea}\email{mariusmillea@gmail.com}
\affiliation{Department of Physics \& Astronomy, University of California, One Shields Avenue, Davis, CA 95616, USA}

\author{Emmanuel Schaan}\email{eschaan@stanford.edu}
\affiliation{Kavli Institute for Particle Astrophysics and Cosmology, 382 Via Pueblo Mall Stanford, CA 94305-4060, USA}
\affiliation{SLAC National Accelerator Laboratory 2575 Sand Hill Road Menlo Park, California 94025, USA}

\begin{abstract}

Future low-noise cosmic microwave background (CMB) lensing measurements from e.g., CMB-S4 will be polarization dominated, rather than temperature dominated.
In this new regime, statistically optimal lensing reconstructions outperform the standard quadratic estimator, but their sensitivity to extragalactic polarized foregrounds has not been quantified.
Using realistic simulations of polarized radio and infrared point sources, we show for the first time that optimal Bayesian lensing from a CMB-S4-like experiment is insensitive to the expected level of polarized extragalactic foregrounds 
after
masking,
as long as an accurate foreground power spectrum is included in the analysis.
For more futuristic experiments where these foregrounds could cause a detectable bias, we propose a new method to jointly fit for lensing and the Poisson foregrounds, generalizing the bias hardening from the standard quadratic estimator to Bayesian lensing.

\end{abstract}

\maketitle

\section{Introduction}

Many properties of the large-scale structure in the Universe are imprinted on the cosmic microwave background (CMB) photons, as they travel from the surface of last scattering towards us.
In particular, gravitational lensing causes arcminute distortions in the CMB, coherent on degree scales \cite{Lewis06, Hanson10},
which have been measured with high precision by experiments like the Wilkinson Microwave Anisotropy Probe (\textit{WMAP}) \cite{Hinshaw13, Smith07}, \textit{Planck} \cite{Planck18,Carron_2022}, the Atacama Cosmology Telescope (ACT) \cite{Fowler07, Swetz11, Thornton16, Henderson16,Das2011,sherwin2011,PhysRevD.95.123529,ACT:2023dou,ACT:2023kun}, the South Pole Telescope (SPT) \cite{Benson14,van_Engelen_2012,Story_2015,Omori_2017,Wu_2019,Bianchini_2020,Millea_2021,pan2023measurement} and POLARBEAR \cite{Inoue16,polarbear} .
Future CMB lensing measurements from Simons Observatory (SO) \cite{SO19}, CMB-S4 \cite{CMBS419} and CMB-HD \cite{Sehgal20} promise to reveal the masses of the neutrinos, improve the detectability of primordial gravitational waves, and measure the mass of galaxy clusters too distant to weigh with galaxy lensing.

Extragalactic foregrounds, such as the thermal and kinematic Sunyaev-Zel'dovich effects (tSZ and kSZ), the cosmic infrared background (CIB) and radio galaxies, are known to be major contaminants to CMB temperature maps, on the small scales where CMB lensing is reconstructed.
They produce a dominant bias to CMB lensing quadratic estimators (QE) \cite{VanEngelen14, Osborne14, Ferraro18, Schaan19, Baxter19, Sailer20} for ACT, SPT and Simons Observatory.
This can be successfully mitigated via a combination of masking, multi-frequency cleaning including gradient-cleaned estimators \cite{Madhavacheril18, Darwish20}, and modified estimators like the shear-only estimator \cite{Schaan19,PhysRevD.108.063518,carron2024spherical} and the bias-hardened quadratic estimators \cite{Osborne14, Namikawa13, plancklensing2013, Sailer20, Sailer21, Sailer23, Darwish23}, which leverage the differing spatial symmetries of the lensing deflection and the foregrounds.

However, the extremely high sensitivity of CMB-S4 brings CMB lensing measurements into a qualitatively new regime.
First, the lensing reconstruction is dominated by polarization, where the extragalactic foregrounds are different.
Second, the standard quadratic lensing estimator is suboptimal compared to iterative and Bayesian lensing techniques \cite{Carron17, Millea19, Carron19, Millea20,Legrand_2022,PhysRevD.108.103516}.
In this paper, we compute for the first time the response of Bayesian lensing to extragalactic foregrounds. We focus on CMB polarization exclusively, since this is the regime where Bayesian lensing improves the lensing signal-to-noise (SNR) over the QE most.

Previous work suggests that polarized extragalactic foregrounds do not bias the QE significantly \cite{Smith09} (see the bottom panel of their Fig.~13)\footnote{However, they only consider radio point sources, assuming that infrared point sources would have a smaller or comparable effect.}.
For the QE, the lensing bias from a given foreground is entirely determined by the foreground's trispectrum and its bispectrum with the true CMB lensing convergence.
However, Bayesian methods extract information from arbitrarily high $n$-point functions of the CMB map, and may thus be sensitive to higher-point functions of the foregrounds.
Predicting the effect of foregrounds on Bayesian lensing is therefore non-trivial. It is also a crucial question to answer, in order to assess the benefit of Bayesian lensing over the QE for CMB-S4. Previous works have studied the impact of foregrounds in both temperature and polarization on maximum likelihood cluster lensing \citep{Raghunathan_2017} or on the impact of Milky Way foreground on iterative estimator \citep{Belkner_2024}. Here for the first time, we quantify the effect of extragalactic polarised foreground on Bayesian CMB lensing.

In this paper, we use simulated source catalogs from \cite{Sehgal10}, along with estimated source polarization fractions from the literature \cite{Lagache19, Bethermin12, Datta19, deZotti05, Gupta20, Puglisi18, Sadler06, Sadler06, Sadler08, Waldram07}, to generate mock polarized foreground maps.
We then perform a Bayesian lensing reconstruction including these mock maps and quantify the resulting bias to CMB lensing.
Finally, looking ahead at futuristic CMB lensing experiments, we outline a new method to jointly reconstruct lensing and fit for the extragalactic foregrounds, relying on the assumption of Poisson-distributed sources.

\section{Simulated polarized extragalactic foregrounds}

On the small scales relevant to CMB lensing reconstruction, the main polarized extragalactic foregrounds are radio galaxies and dusty star-forming galaxies.
Below, we review what is known about them, and describe the process to generate simulations that contain these foregrounds.

\subsection{Radio galaxies}

As reviewed in detail in \cite{Lagache19}, much is known about the flux distribution, spectral energy distribution and nature of radio sources at low frequencies $\nu \lesssim 10$~GHz.
These sources are active galactic nuclei and can be split into two populations, depending on the steepness of their spectra $S(\nu) \propto \nu^\alpha$ at frequencies $\nu \lesssim 10$~GHz.
The steep-spectrum sources, such as BL Lacertae objects, have $\alpha < -0.5$ and are thus less important at CMB frequencies $\sim 100$~GHz than the flat-spectrum or inverted-spectrum sources such as blazars, for which $\alpha \geq 0.5$.
However, extrapolation from low frequencies to the higher CMB frequencies is inexact, as the spectral energy distributions of both steep and flat-spectrum sources deviate from a power law, showing a downturn near the CMB frequencies.
Fortunately, measurements of radio sources at CMB frequencies are available from \textit{Planck} \cite{Planck11VII, Planck11XIII}, ACT \cite{Marriage11, Marsden14, Datta19} and SPT \cite{Mocanu13b, Gupta20}.
Ref.~\cite{Gupta20} finds a root-mean-squared polarization fraction of 2.6\% at 95~GHz and 150~GHz using SPTPol data. 
Similarly, Ref.~\cite{Datta19} finds a mean polarization fraction of 3\%, independent of source brightness, using ACTPol data at 150~GHz.
They predict that for a 1~$\mu$K$\cdot$arcmin sensitivity, the polarized radio source signal at 150~GHz is below the EE noise out to $\ell=6000$.
They find no dependence of the polarization fraction with source flux, consistent with \cite{Puglisi18}.
Ref~\cite{Puglisi18} further finds a slight increase in the polarization fraction as a function of observing frequency.

Furthermore, \cite{Lagache19} points out that the flux distribution of radio sources roughly scales as the inverse flux squared, $n(S) \propto S^{-2}$. 
As a result, the shot noise power spectrum from undetected sources $\propto \int^{S_\text{cut}}_0 dS\ n(S) S^2$ simply scales as the flux detection threshold $S_\text{cut}$.
Thus the shot noise power spectrum from radio sources roughly scales as the experiment sensitivity.
In other words, as the detector noise is reduced, the shot noise from radio sources goes down too, a very encouraging result for a CMB-S4-like experiment.
This is even more encouraging for CMB lensing, where the foreground biases scale as the source bispectrum and trispectrum.
Indeed, the shot noise bispectrum and trispectrum scale as $S_\text{cut}^2$ and $S_\text{cut}^3$ respectively, making them less and less important as the sensitivity improves.

\subsection{Dusty star-forming galaxies}

The thermal emission from dust grains in star-forming galaxies produces the cosmic infrared background (CIB).
Because dust grains can be aligned with the local magnetic field inside a galaxy, this thermal emission can be polarized.
The resulting polarized component of the CIB is the other main foreground in CMB polarization.
Although less is known about the polarization fraction of dusty star-forming galaxies, it is expected to be small, due to the complex magnetic field structure inside galaxies, resulting in an averaging of the polarized emission along the line of sight and within the beam.
Ref~\cite{Gupta20} gives an upper limit of 5\% (resp. 13\%) at 150~GHz (resp. 95~GHz), with no evidence for a variation of the polarized fraction with intensity or observing frequency.
As reviewed in \cite{Lagache19}, measurements for individual galaxies give typical polarization fractions of about 0.4-4\% at 217-353~GHz.
We follow \cite{Lagache19} in assuming a typical polarization fraction of 1\% for dusty star-forming galaxies.
The population of dusty star-forming galaxies is well described by two modes, the so-called star-forming main sequence and the starburst mode \cite{Bethermin12}. The resulting luminosity function is different from that of radio sources, such that the shot noise power spectrum is much less sensitive to the flux cut used for masking \cite{Lagache19}.
Dusty star-forming galaxies are clustered, with a power spectrum described by a 2-halo, 1-halo and shot noise terms \cite{Penin12}. 
The polarized power spectrum, however, is not obtained by simply multiplying the temperature power spectrum by the mean squared polarization fraction \cite{Lagache19}.
Indeed, the polarized 2-halo term vanishes due to the decorrelation of polarization orientations across galaxies, and the 1-halo term is reduced due to the averaging of polarization orientations of galaxies within a given halo.

\subsection{Realistic simulated maps from the Sehgal catalogs} \label{sec: simulation}

\subsubsection{Simulated catalogs}

We generate mock polarized foreground maps, based on the realistic simulated radio and infrared galaxy catalogs from \cite{Sehgal10}\footnote{The catalogs are publicly available at \url{https://lambda.gsfc.nasa.gov/simulation/tb_sim_ov.cfm}}.

In this simulation, dark matter halos from an N-body simulation are populated with infrared galaxies, following the model of \cite{Righi08}.
In all these infrared galaxies, the dust spectral energy distribution is assumed to be identical, a modified blackbody with power law index $\beta=1.4$, and a dust temperature $T_0=25$~K at $z=0$, evolving with redshift according to \cite{Blain99}.
The infrared galaxies have the same infrared luminosity $L_\star = 1.25\times 10^{12} L_\odot$, but their number within each halo depends on the halo mass and redshift.
Specifically, halos with mass less than $M_1 = 2.5\times 10^{11} M_\odot$ do not contain any IR galaxies, whereas the halo occupation number for more massive halos is equal to $M/M_2\ e^{-M/M_\text{cool}} W_\text{sfr}(z)$, where $M$ is the halo mass, $M_2=3\times 10^{13} M_\odot$, $M_\text{cool} = 5\times 10^{14} M_\odot$ accounts for the longer cooling time in more massive halos, and $W_\text{sfr}(z)$ models the mean evolution of the star formation rate with redshift.
These parameters were chosen to match observations of the mean CIB \cite{Fixsen98} and its observed spectral slope \cite{Knox04}, the source counts from \cite{Coppin06} and the rough shape of the CIB power spectrum observed in \cite{Lagache07, Viero09, Reichardt09}.
Halos with mass above $10^{13} M_\odot$ are resolved in the simulations.
As a result, the simulated galaxies inside these halos have a realistic clustering.
Halos with lower masses are added artificially, by Poisson sampling the halo mass function from \cite{Jenkins01} and placing them according to a Gaussian spatial distribution with width 10~Mpc around the more massive halos.
The galaxies inside the lower mass halos thus have some amount of clustering.

The emission from the radio sources in \cite{Sehgal10} is modelled as the sum of an AGN core and AGN lobes, whose spectral energy distributions differ and follow \cite{Lin09}.
The distribution of AGN luminosities is assumed to follow a broken power law, and their halo occupation number is modelled as a power law with halo mass.
The parameters are adjusted to fit the observed radio luminosity function in Laing83.
The AGNs are split into two populations with different redshift evolutions, following \cite{Wilman08}.
The model parameters are then selected to match the observed source counts in \cite{deZotti05, Bondi03, Huynh05, Cleary05, Wright09, Reichardt09}.
 
\begin{figure}
    \centering
    \includegraphics[width=0.8\linewidth]{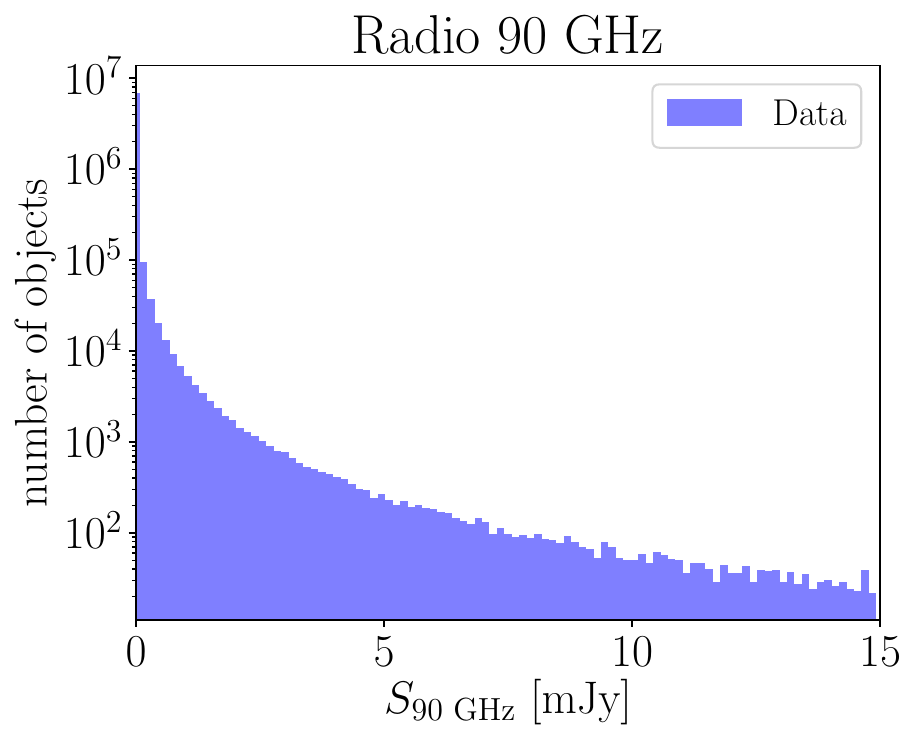} \\ 
    \vspace{-5pt} 
    \includegraphics[width=0.8\linewidth]{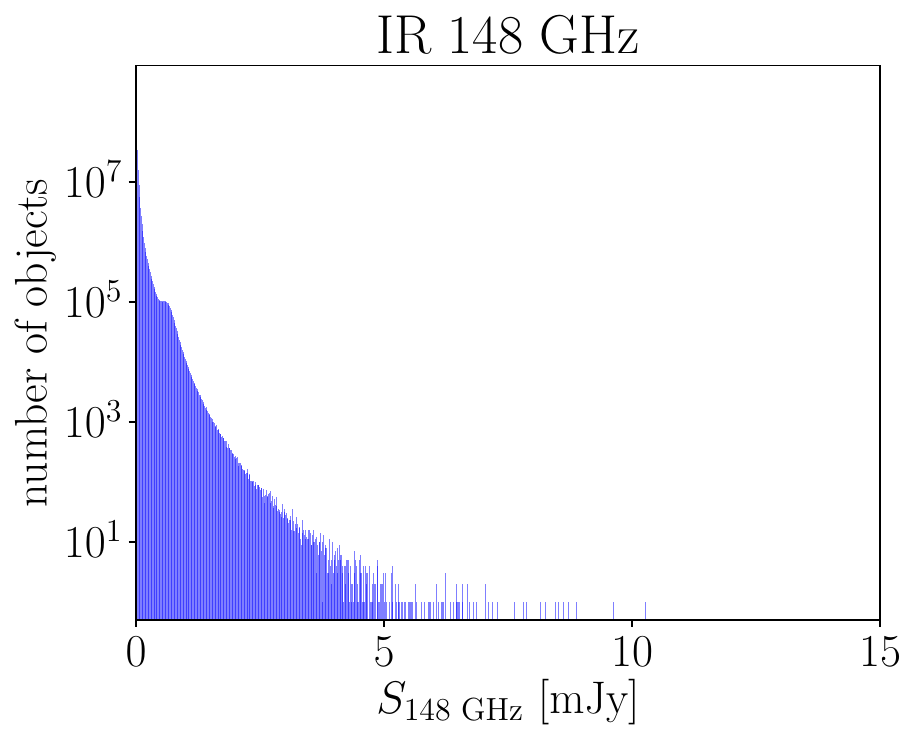} \\ 
    \caption{Flux distribution of the radio (top) and infrared (bottom) sources in the \cite{Sehgal10} catalogs. 
The y-axis is the number of galaxies in the octant of the sky within each flux bin.
In the analysis, we will be masking sources with fluxes above 2, 5 and 10mJy.}
    \label{fig:radio_ir_hist}
\end{figure}

The flux distributions of the radio and IR sources are shown in Fig.~\ref{fig:radio_ir_hist}, and compared with the expected flux cuts for CMB-S4.

\subsubsection{Simulated polarized maps}

To convert these galaxy catalogs into polarized maps, we assume a fixed polarization fraction of 3\% and 1\% for radio galaxies and infrared galaxies, respectively.
In reality, the polarization fraction is expected to vary from galaxy to galaxy. 
This would modify the amplitude of the $n$-point functions of the polarized map, and the ratio of $m$ to $n$-point functions.
However, for an unclustered population of sources, this does not add any scale dependence to the $n$-point functions.
In practice, the lensing reconstruction is dominated by the small scales where this approximation is valid.
Since the polarization angle for each source is determined by the detail of the local interstellar medium physics, we assume that distinct galaxies have independent polarization angles.
As a code check, we verified that our code reproduces the publicly available maps from \cite{Sehgal10} when generating maps of the unpolarized intensity.

\subsubsection{Flat square cutouts}

Since the catalogs (and maps) from \cite{Sehgal10} only cover one octant of the sky, and to simplify the lensing analysis, we extract 40 non-overlapping flat square cutouts, 10 degrees on a side, from this octant.

\begin{figure}
    \centering
    \includegraphics[width=0.8\linewidth]{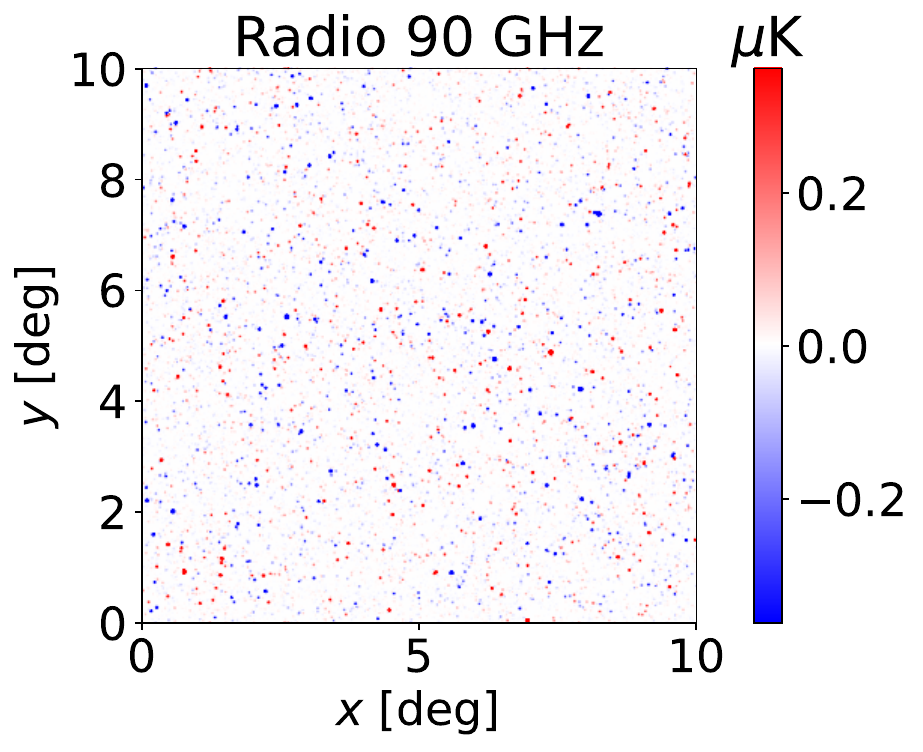} \\ 
    \vspace{-5pt} 
    \includegraphics[width=0.8\linewidth]{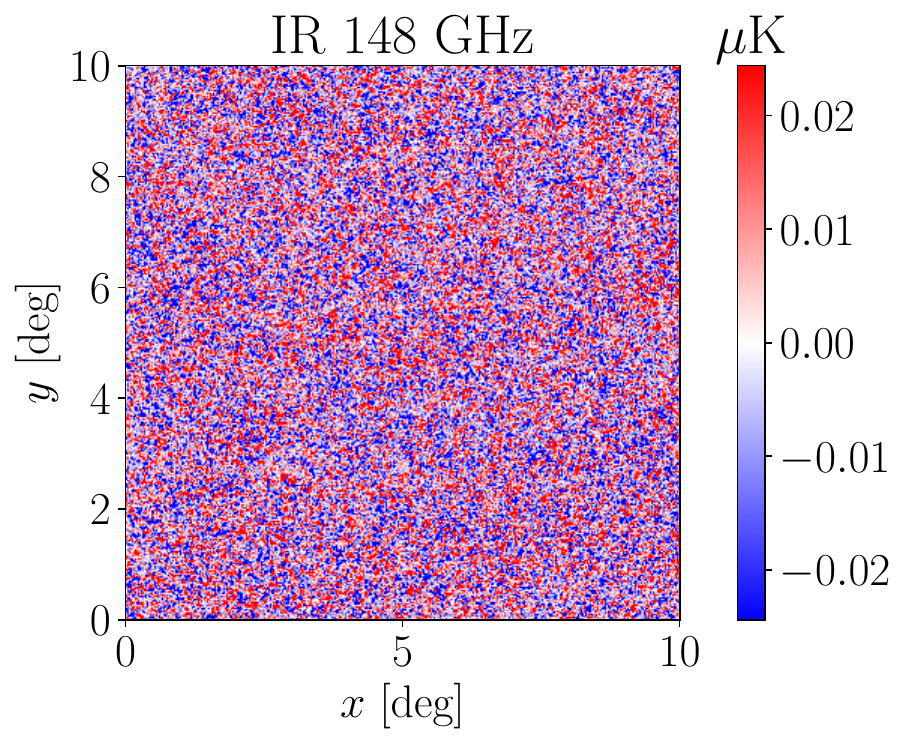} \\ 
    \vspace{-5pt} 
    
    \includegraphics[width=0.8\linewidth]{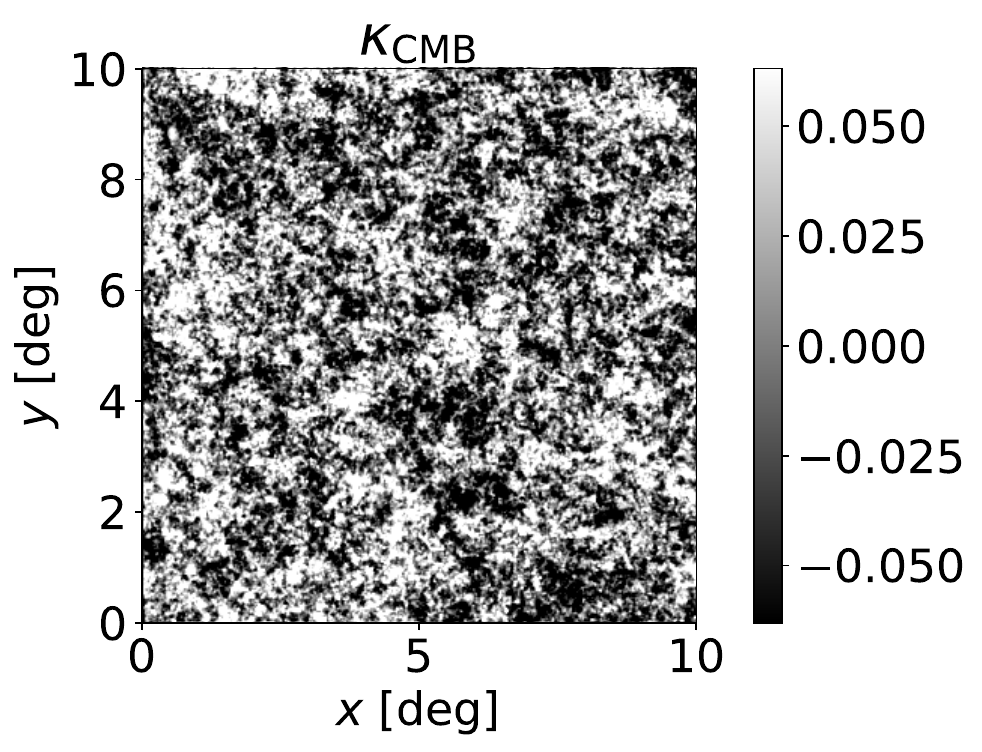} 
    \caption{Cutouts from the simulated Q maps of the polarized emission from radio (top panel) and infrared (middle panel). To make the radio point sources more easily visible, these maps are shown convolved with a Gaussian beam with FWHM$=2'$ with no masking applied. The U maps, not shown, are statistically identical. The bottom cutout shows the lensing convergence cutout on the same patch.}
    \label{fig:radio_ir_maps}
\end{figure}

\subsubsection{Point source masking}

In this simulated analysis, we mimic a template subtraction of the point sources, detected at either the map level or the time-ordered data (TOD) level, rather than masking the observed temperature map and inpainting the masked areas.
This procedure avoids the bias from preferentially removing parts of the maps that are in high lensing convergence regions, since they are in high foreground regions~\cite{PhysRevD.106.023525,PhysRevD.103.043535}.

We generate point source masks for each foreground cutout as follows.
We match-filter each temperature map cutout, assuming a point source profile and a noise power spectrum equal to the total temperature power spectrum (lensed CMB plus all foregrounds at the given temperature, plus detector noise).
We then mask any pixel above a flux threshold, e.g., 2, 5 and 10 mJy.
We then paint a $3'$ radius disk around each masked pixel in the point source maps to produce the final point source mask.
The point source mask is then applied to the foreground-only map, before adding it to the lensed CMB and detector noise.
Hence the lensed CMB map is not itself masked.

In particular, we assume that the point sources are detected using temperature only, a conservative choice. 
While the source signal is smaller in polarization by a factor $\sim 100$, the confusion noise from the CMB is also smaller there, such that polarization data may contain additional useful information on the point sources.

\subsubsection{Polarized foreground power spectra}

\begin{figure}
\centering
\includegraphics[width=0.9\columnwidth]{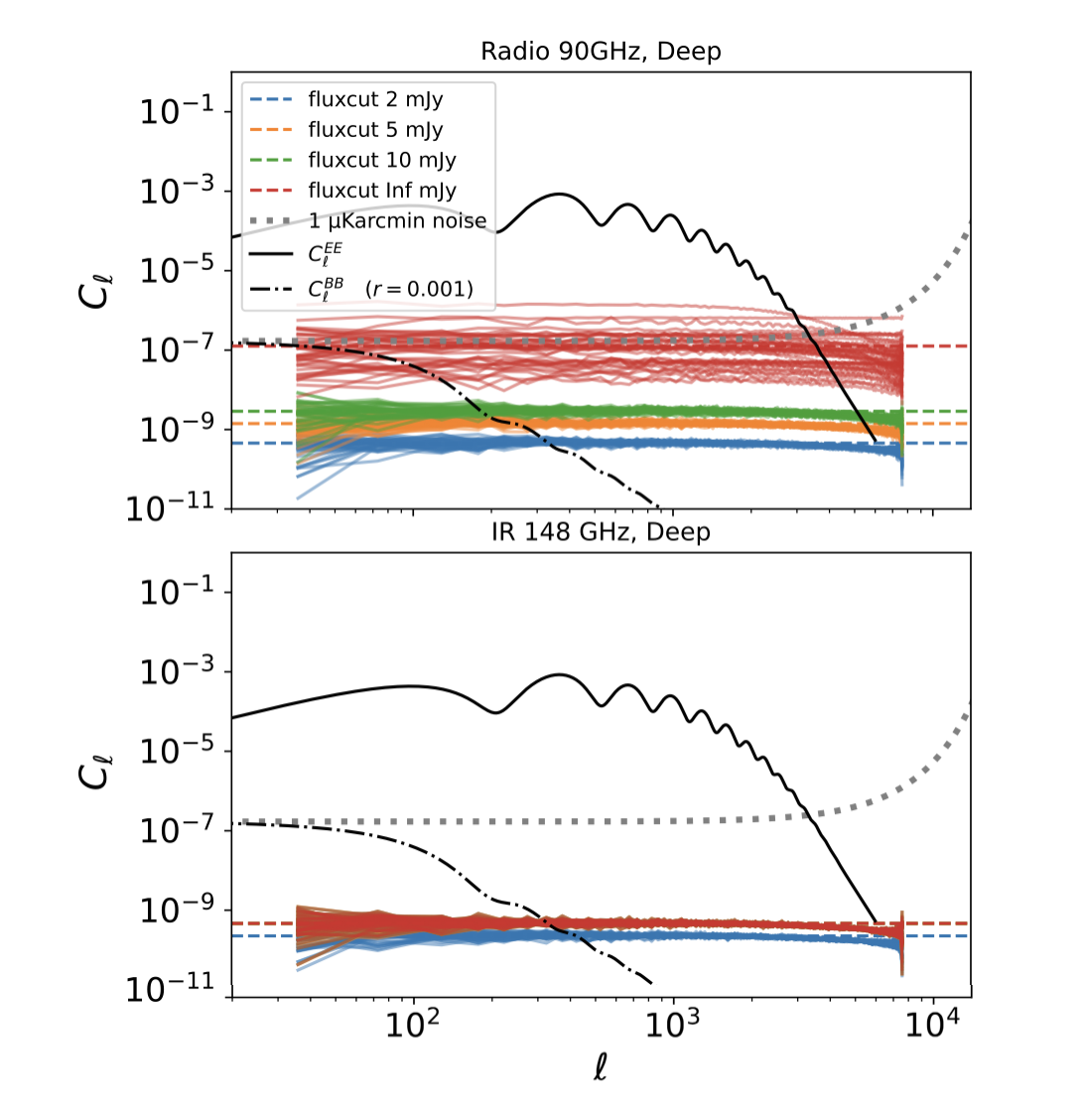}
 \caption{
The $QQ$ power spectrum of foreground simulations for the deep configurations is described in Table \ref{tab:cmbs4_specs} (wide configurations have similar levels of foreground power, with most cases differing from the deep configurations by $\sim5\%$). The top subplot corresponds to radio sources and the bottom plot for infrared sources.
Different colors represent different flux cuts in mJy with red corresponding to no masking. 
Each line corresponds to the power spectra measured in each of the non-overlapping $10^\circ\times10^\circ$ cutouts. 
For comparison, solid black lines show lensed scalar $EE$ and $BB$ spectra and the dotted grey line corresponds to $1\,{\rm \mu K\,arcmin}$ noise. }
\label{fig:radio_ir_power}
\end{figure}

The power spectra of the 40 cutouts are shown in Fig.~\ref{fig:radio_ir_power}, as a function of the point source mask threshold used.
The power spectrum of polarized IR sources is negligible compared to the detector white noise on all scales, even in the absence of masking. 
It is also smaller than the lensed scalar E and B power spectra out to $\ell\sim 6000$.
On the other hand, the unmasked power spectrum of polarized radio sources can be as much as ten times larger than the detector noise.
The power spectrum scatter across patches is also very large. 
This scatter is in part non-Gaussian, sourced by the Poisson fluctuations in the rare brightest sources in the maps. 
Masking significantly reduces the radio source power spectrum well below the detector noise, as anticipated above.
Masking the bright sources also significantly reduces the power spectrum scatter across patches.
Indeed, it leaves only the more numerous fainter sources, thus reducing the non-Gaussianity of the maps, since Poisson fluctuations in a larger mean number of sources are closer to Gaussian.

\begin{figure*}
    \centering
    {\includegraphics[width=\linewidth]{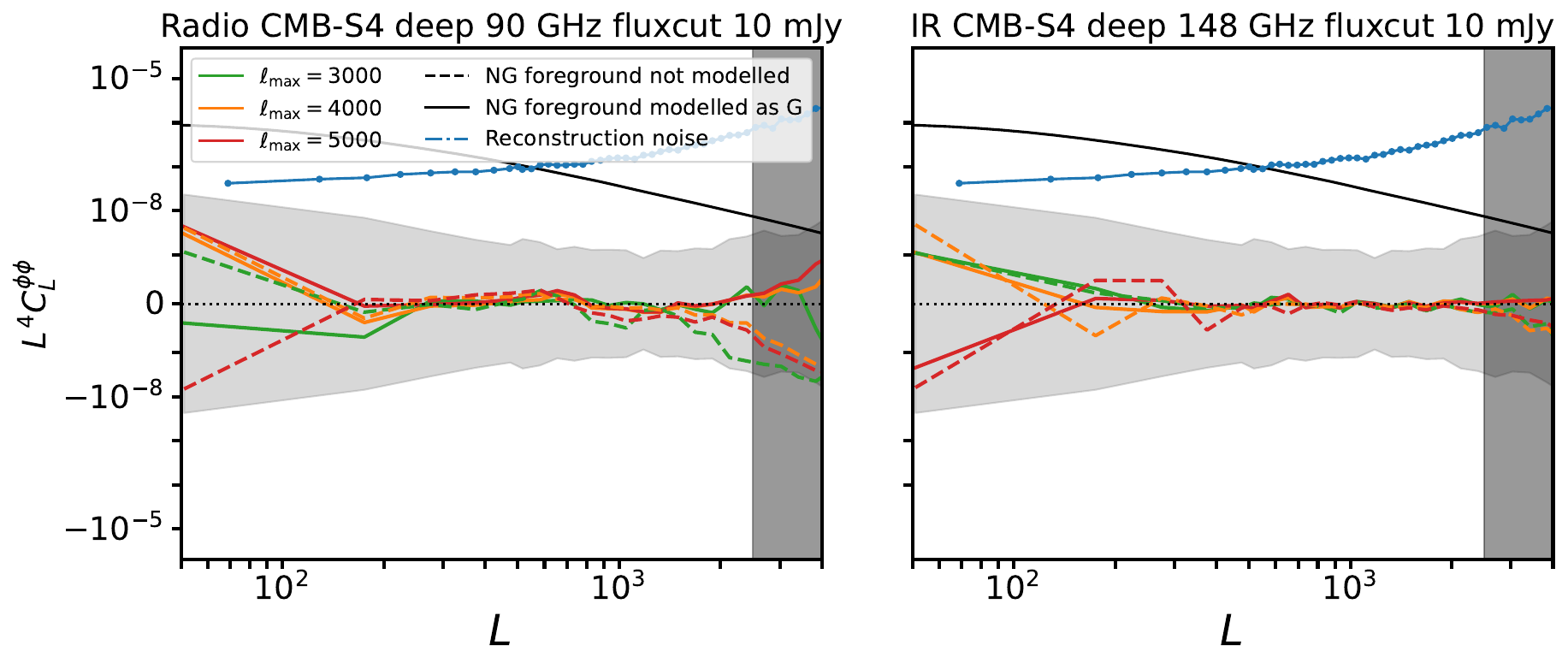}}
    \caption{Lensing bias for radio sources (left panel) and infrared sources (right panel) with S4-deep configuration, fixed mask fluxcut at $10$mJ and varying $\ell_\mathrm{max}$ cut. The shaded grey region shows the $1-\sigma$ uncertainty for the lensing reconstruction with  $\ell_\mathrm{max}=3000$ and the corresponding reconstruction noise is in blue. The dashed curves show the bias induced in lensing when the simulations used in the modelling $d^\prime$ do not have a foreground contribution, giving a biased results for all the $\ell_\mathrm{max}$ used. The solid lines shows that foreground biases are effectively mitigated when the simulations used incorporate a Gaussian foreground with a power spectrum matching that of the data. }
    \label{fig.lensing_bias_diagram}
\end{figure*}

The corresponding equivalent white noise levels of the masked radio and IR sources are shown in Table~\ref{tab:radio_ir_equivalent_noise}.
\begin{table}
    \setlength{\tabcolsep}{7pt}
    \renewcommand{\arraystretch}{1.2}
    \begin{tabular}{l|ccccc}
        \toprule
        {}             & Radio    & Radio & Radio & Radio & IR       \\
        $\Scut$:       & $\infty$ & 10\,mJy    & 5\,mJy     & 2\,mJy     & $\infty$ \\
        \midrule
        150\,GHz       &  0.63    &  0.10 & 0.07  & 0.04  & 0.07     \\
        90\,GHz        &  1.46    &  0.25 & 0.18  & 0.11  &          \\
        \bottomrule
    \end{tabular}
    \caption{Mean effective white-noise level in $Q$ in $\mu$K.arcmin for radio and IR sources masked based on their flux in temperature, with the flux cut, $\Scut$, labeled in each column. The power spectra for individual simulations for these same cases is shown in Fig.~\ref{fig:radio_ir_maps}. Note that in the case of unmasked radio sources, there is a very large amount of non-Gaussianity, meaning individual realizations can have almost an order of magnitude larger or smaller power than the mean.}
    \label{tab:radio_ir_equivalent_noise}
\end{table}

\section{Bias to the lensing reconstruction}

\subsection{CMB-S4 specifications \& Lensing reconstruction}

To quantify the effect of polarized foregrounds on Bayesian CMB lensing from CMB-S4, we proceed as follows.

\subsubsection{Simulated CMB-S4 maps}

We start with the polarized foreground cutouts described above and mask them with a flux threshold of 2, 5, 10 mJy or no mask.
We then generate a Gaussian random field (GRF) for the unlensed polarized CMB.
We lens it with a convergence field $\kappa$ correlated with the foreground.
To do this, we extract a cutout from the simulated convergence map of \cite{Sehgal10} at the same location as the foreground cutout considered.
The simulated convergence map, from N-body simulations, is thus appropriately correlated with the radio and IR sources, and its non-Gaussianity realistically describes the nonlinearity of the matter density field in the Universe.
So that the same lensing algorithm is used for both simulation and analysis, we use the code \texttt{LenseFlow} \cite{PhysRevD.100.023509} to perform the lensing operation on the simulated CMB map.
To simulate an observed CMB-S4 map, we apply a Gaussian beam to the sum of the lensed CMB and foreground and add a GRF to mimic the detector and atmospheric noise.
We assume two different specifications corresponding approximately to S4-Deep and S4-Wide, as shown in Table.~\ref{tab:cmbs4_specs}.
\begin{table}
    \setlength{\tabcolsep}{8pt}
    \renewcommand{\arraystretch}{1.2}
    \begin{tabular}{l|cccc}
        \toprule
        & \multicolumn{2}{c}{S4-Deep} & \multicolumn{2}{c}{S4-Wide} \\
        & 90 & 148 & 90 & 148 \\
        \midrule
        $f_{\rm sky}$ & \multicolumn{2}{c}{3\%} & \multicolumn{2}{c}{40\%} \\
        $\ell_{\rm knee}$ & \multicolumn{2}{c}{200} & \multicolumn{2}{c}{700} \\
        $\alpha_{\rm knee}$ & \multicolumn{2}{c}{2} & \multicolumn{2}{c}{1.4} \\
        Beam FWHM [arcmin] & 2.3 & 1.5 & 2.2 & 1.4 \\
        $EB$ Noise [$\mu$K-arcmin] & 0.68 & 0.96 & 2.9 & 2.8 \\
        $TT$ Noise [$\mu$K-arcmin] & 0.48 & 0.68 & 2.0 & 2.0 \\
        \bottomrule
    \end{tabular}
    \caption{
    CMB-S4 specifications for the deep and wide surveys, following Tables 2.2 and 2.3 in \cite{CMBS419} and the \href{https://cmb-s4.org/wiki/index.php/Expected_Survey_Performance_for_Science_Forecasting}{CMB-S4 public wiki page}. We also assume that in polarization the 1/f atmospheric noise is subdominant and scales with white noise with a fixed $\ell_\mathrm{knee}=200$ and $\ell_\mathrm{knee}=700$ for the deep and wide settings respectively. We adopt for the lensing reconstruction $\ell_\mathrm{max}=3000,4000,5000$.
    }
    \label{tab:cmbs4_specs}
\end{table}

In the simulated maps, the masking operation was applied only to the foreground maps, rather than the final observed map.
This approach mimics the point source subtraction or inpainting performed at the time-ordered data level in ground-based experiments.
This leaves no holes in the map, thus avoiding a complex point source mask and the associated lensing mean field.

\subsubsection{Bayesian lensing reconstruction}
\label{sec:bayesian_lensing}

We quantify foreground biases to an optimal lensing analysis by performing Bayesian inference of the lensing potential power spectrum, $C_L^{\phi\phi}$. This involves working with the marginal posterior probability function $\mathcal{P}(C_L^{\phi\phi}\,|\,d)$, where $d$ are the simulated data. The marginal posterior is obtained by integrating the joint posterior probability function over the lensing potential field $\phi$ and the unlensed CMB Q/U polarization fields, $f$, 
\begin{equation}
    \mathcal{P}(C_L^{\phi\phi}\,|\,d) = \int {\rm d}\phi \, {\rm d}f\, \mathcal{P}(f,\phi,C_L^{\phi\phi}\,|\,d).
    \label{eq:margpost}
\end{equation}
The joint posterior models the simulated data described above, and in this case is: 
\begin{widetext}
    \begin{align}
    \label{eq:jointposterior}
    \mathcal{P}\big(\,f,\phi,\Clpp\,|\,d\,\big) \;\propto 
    \frac{\exp\left\{ -\cfrac{\left(d - \op{M} \, \op{B} \, \Len[\phi] \, f \right)^2}{2 \,(\Cn + \Cg)} \right\}}{\det (\Cn+\Cg)^{\nicefrac{1}{2}}} \;
    \frac{\exp\left\{ -\cfrac{f^2}{2\,\Cf} \right\}}{\det  \Cf^{\nicefrac{1}{2}}} \;
    \frac{\exp\left\{ -\cfrac{\phi^2}{2\,\Cphi[\Clpp]} \right\}}{\det  \Cphi[\Clpp]^{\nicefrac{1}{2}}},
    \end{align}
\end{widetext}
where $\op{M}$ is a Fourier mask picking out modes between $\ell_{\rm min}\,{<}\,\ell\,{<}\,\ell_{\rm max}$, $\op{B}$ are the beams, $\Len[\phi]$ is the lensing operation, and $\Cn$, $\Cf$, $\Cg$, and $\Cphi[\Clpp]$ are covariances for the noise, the unlensed CMB, foregrounds, and for $\phi$, with the latter explicitly depending on the given \Clpp. We use the shorthand $x^2/\op N$ to represent $x^\dagger \op N^{-1} x$. In cases where we consider modeling the foregrounds as Gaussian, their power spectrum is included as $\Cg$, or in cases where they are considered as unmodeled, this term is set to zero. 

To perform inference, we use the MUSE algorithm, which provides and approximate best-fit to the marginal posterior, $\widehat C_L^{\phi\phi}$, and the posterior covariance matrix $\Sigma_{LL^\prime}$. For CMB lensing inference, these approximations were shown to be highly accurate in \cite{PhysRevD.105.103531}. 

MUSE iteratively estimates the best-fit, $\widehat C_L^{\phi\phi}$ by taking Newton–Raphson steps of the form, 
\begin{equation}
    C_L^{\phi\phi} \rightarrow C_L^{\phi\phi} - K_{LL^\prime}^{-1} \, g_{L^\prime}
\end{equation}
starting from an initial guess, which here we take as the lensing power spectrum of the input simulations. The gradient direction, $g_{L}$ is given by
\begin{multline}
    g_L = \frac{d}{d(\Clpp)} \mathcal{P}(\hat f_d, \hat \phi_d, \Clpp\,|\,d) \\ - \left\langle \frac{d}{d(\Clpp)} \mathcal{P}(\hat f_{d^\prime}, \hat \phi_{d^\prime}, \Clpp\,|\,d^\prime) \right\rangle_{d^\prime \sim \mathcal{P}(d^\prime\,|\,\Clpp)}.
    \label{eq:gradient}
\end{multline}
where $\hat f_d$ and $\hat \phi_d$ are joint maximum a posteriori (MAP) estimates of the unlensed CMB and lensing potential fields,
\begin{equation}
    \hat f_d, \hat \phi_d = \underset{f,\phi}{\rm argmax} \; \mathcal{P}(f,\phi,\Clpp\,|\,d),
\end{equation}
with the subscript $d$ indicating the dependence of the MAP on the data and the dependence on \Clpp left as implied.
The covariance, $\Sigma_{LL^\prime}$, is computed via gradients and Jacobians of the joint posterior evaluated at the MAP akin to Eqn.~\eqref{eq:gradient}, and can be found in \cite{millea2020}.

The $K_{LL^\prime}$ matrix in the Newton-Rhapson step can in general be anything but for fastest convergence should match the Jacobian of Eqn.~\eqref{eq:gradient}. As a simplification, we have noted that when using the MUSE covariance itself ($K_{LL^\prime}^{-1}=\Sigma_{LL^\prime}$), MUSE converges to a satisfactory solution in only a single iteration up to $L\,{\sim}\,2500$, high enough to capture all appreciable signal. This is verified in Appendix~\ref{app:convergence_steps} by inputting a know deviation to $\Clpp$ and confirming that it is recovered in a single iteration. 

With these ingredients in place, we can quantify foreground bias by computing the one-iteration MUSE $\widehat C_L^{\phi\phi}$ estimate, where the $N$-body simulated data are input as $d$ in Eqn.~\eqref{eq:gradient}, and then averaging this $\widehat C_L^{\phi\phi}$ over several such simulations. Note that in this average, the term in angle brackets in Eqn.~\eqref{eq:gradient} is the same as it does not depend on $d$ but rather on simulations from the assumed analysis model, $d^\prime \sim \mathcal{P}(d^\prime\,|\,\Clpp)$, so does not need to be recomputed for each realization. Thus the bias calculation takes the convenient form of computing the average difference in gradients at the MAP for a set of simulations based on realistic foregrounds and a set of simulations from the analysis model, where the analysis model assumes either no foreground at all or foregrounds modeled as Gaussian with some power spectrum,

\begin{multline}
    \Delta \widehat  C_{L, \rm bias}^{\phi\phi} = \\ -K_{LL^\prime}^{-1} \left(\left\langle \frac{d}{d(\Clpp)} \mathcal{P}(\hat f_{d}, \hat \phi_{d}, \Clpp\,|\,d) \right\rangle_{d \sim \{\rm realistic \; sims\}}\right. \\ - \left. \left\langle \frac{d}{d(\Clpp)} \mathcal{P}(\hat f_{d^\prime}, \hat \phi_{d^\prime}, \Clpp\,|\,d^\prime) \right\rangle_{d^\prime \sim \mathcal{P}(d^\prime\,|\,\Clpp)}\right).
    \label{eq:bias}
\end{multline}

We note that when comparing these biases to uncertainties throughout the rest of the paper, we take the uncertainties from $\Sigma_{LL^\prime}$ computed in the absence of any foregrounds.
Indeed, polarized foregrounds have a negligible impact on the lensing reconstruction noise, since their power spectrum is a negligible contribution to the total map power spectrum for E and B-modes (Fig.~\ref{fig:radio_ir_power}).
On the other hand, their power spectrum is not negligible compared to the lensing correction to the E and B power spectra: they can thus cause a significant bias.

Our lensing reconstruction makes several assumptions, which we highlight here.
Our analysis assumes a flat sky rather than a curved sky.
This is an appropriate approximation since most of the lensing information comes from small angular scales and the size of the cut-outs are small. 

While the simulated unlensed CMB and the detector noise are generated as continuous periodic GRFs on each cutout, the foregrounds and the lensing convergence are not necessarily periodic, as they were extracted from a full-sky map. 
This could in principle introduce a lensing mean field from the cutout footprint.
In practice, we find this not to be the case: in the absence of foregrounds, we recover an unbiased lensing power spectrum.
This can be understood, since the foregrounds are (non-Gaussian) white noises, i.e. each pixel value is independent from the others. 
As a result, the question of periodicity on the scale of the cutout is irrelevant. 
The lensing convergence field is not white, but still has most of its power on small scales, making it similar to the white noise case.
Furthermore, \cite{Namikawa14} shows that for the lensing QE, the odd parity of the EB lensing estimator implies that its mean field has no response to the large-scale ($\ell\sim 0$) modes of the mask. While this is not the case for the EE estimator, most of the lensing signal-to-noise comes from EB. We suspect that a similar phenomenon could be occurring in Bayesian lensing.

\subsection{Intuition: Point sources and dipole lensing bias} 
\label{sec.psdp}

Before quantifying the effect of foregrounds on the reconstructed lensing power spectrum in \S\ref{sec:bayesian_lensing}, we describe here the qualitative intuition for the impact of polarized sources on the lensing reconstruction.
We contrast the case of Bayesian lensing with that of the standard QE.

\subsubsection{Bayesian lensing intuition}
\label{sec.intuition_bayesian_lensing}

\begin{figure}[!htb]
    \centering
    \includegraphics[width=0.8\linewidth]{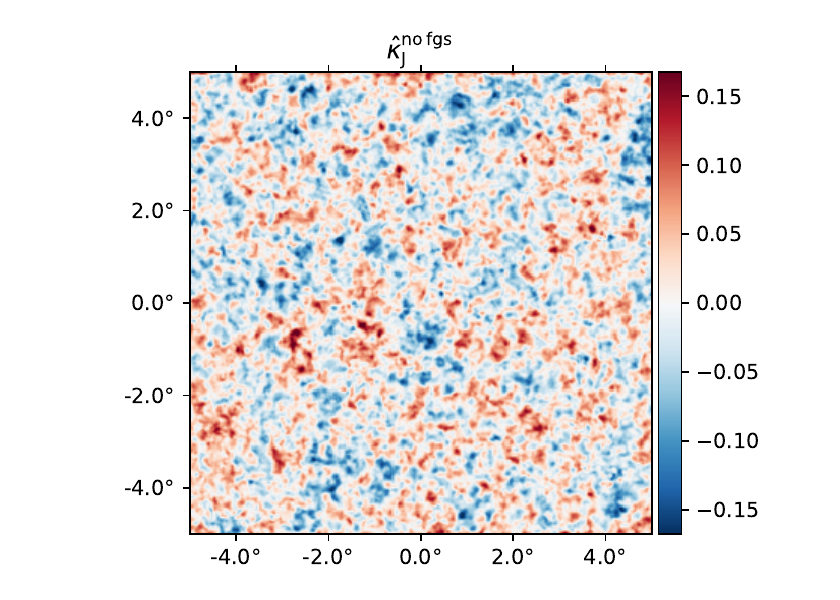} \\ 
    \vspace{-8pt} 
    \includegraphics[width=0.8\linewidth]{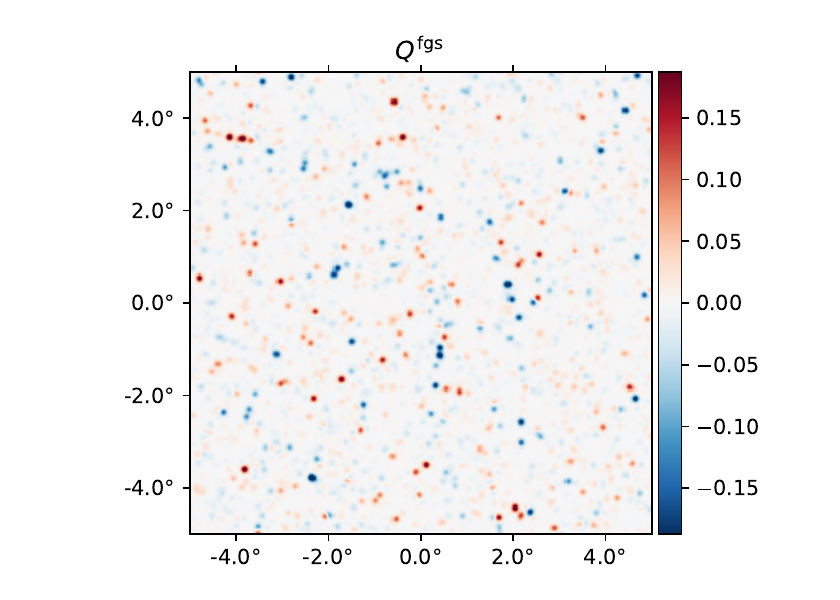} \\ 
    \vspace{-8pt} 
    \includegraphics[width=0.8\linewidth]{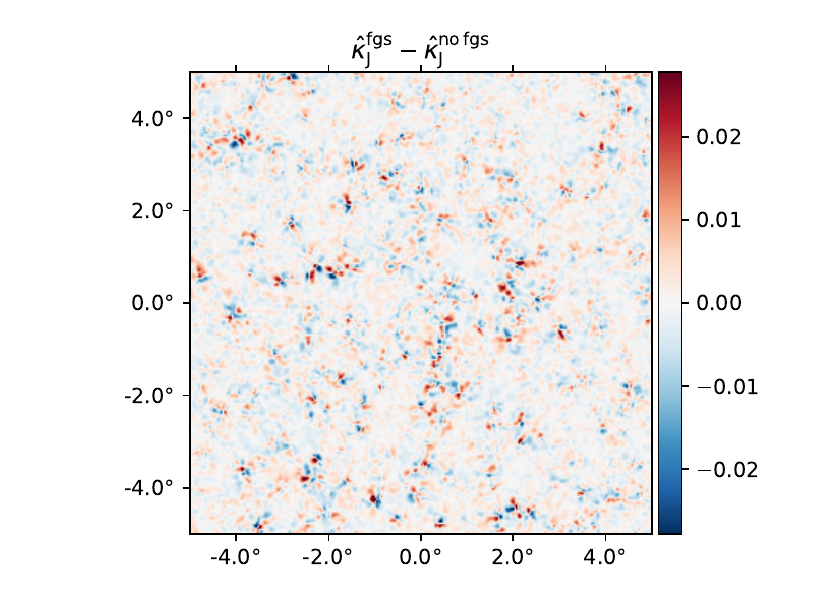} 
    \caption{The reconstructed CMB lensing map in the absence of foregrounds (top panel), acquires a bias (bottom panel) from the polarized radio point sources (central panel).
As the color scales indicates, the typical lensing bias at the map level is about 10\% of the typical lensing signal, consistent with a percent bias in the lensing power spectrum.
As expected, a dipole-like bias in the convergence map is seen at the position of the point sources.
}
    \label{fig:bias_lensing_maps}
\end{figure}

We visually examine the reconstructed convergence maps from Bayesian lensing and QE.
Fig.~\ref{fig:bias_lensing_maps} shows the reconstructed CMB lensing convergence field without foregrounds (top panel), the polarized radio sources in $Q$ (central panel) and the reconstructed convergence field  with foregrounds (bottom panel).
As expected, the reconstructed convergence map acquires a visible bias due to foregrounds, localized at the position of the foreground sources, in the form of a local dipole pattern.

\begin{figure}
\centering
\includegraphics[width=\columnwidth]{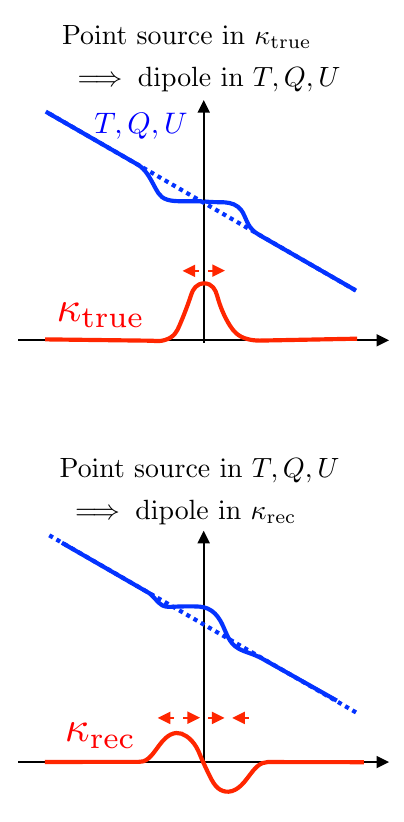}
 \caption{
Top panel: A point source in the convergence map, such as a galaxy cluster, produces a characteristic dipole in the observed CMB maps. This effect can be used to reconstruct lensing on small scales.
Bottom panel: Foregrounds add point sources to the observed CMB maps, rather than the convergence field.
This observed point source can be mimicked by a dipole in the lensing convergence, aligned with the unlensed CMB gradient. In order to explain the foreground sources, the lensing algorithm therefore adds these convergence dipoles at the positions of the sources. This biases the lensing reconstruction.
}
\label{fig:intuition_kappa_dipoles}
\end{figure}

This can be understood intuitively by considering the effect of a localized convergence monopole and dipole on the lensed CMB map, see Fig.~\ref{fig:intuition_kappa_dipoles}.
A point source in the convergence map produces a dipole in the lensed CMB map, aligned with the unlensed CMB gradient ~\cite{Seljak_2000,Hu_2007}(Fig.~\ref{fig:intuition_kappa_dipoles}, top panel).
This is the dominant observable effect of CMB lensing on small scales, and is also the signature that Bayesian lensing methods are relying on to reconstruct lensing. 
The same effect is also used to reconstruct CMB lensing from clusters on small scales ~\cite{Hu_2007}, or in the ``gradient inversion'' (GI) estimator \citep{Horowitz_2019,PhysRevD.100.023547}.
Instead, foregrounds add point sources in the observed CMB map, rather than in the convergence map.
Furthermore, since the foregrounds are correlated with the true lensing convergence, these spurious point sources in the observed CMB map are colocated with corresponding point sources in the convergence map.
The Bayesian lensing model cannot account for these temperature or polarization point sources from the unlensed CMB (which does not have power on such small scales) nor from the Gaussian noise and foreground model.
It therefore fits these point sources by modifying the reconstructed convergence field.
As shown in the bottom panel of Fig.~\ref{fig:intuition_kappa_dipoles}, a dipole in the convergence field, aligned with the unlensed CMB gradient, produces a point source-like structure in the lensed CMB maps.
The lensing algorithm thus adds these spurious convergence dipole patterns at the positions of the foreground sources, in order to explain away the foreground.
These spurious dipoles in the convergence map enhance the recovered lensing power spectrum, producing a positive bias to 
$C_L^{\phi\phi}$.

Visually, the polarized point sources do not appear to cause a large-scale bias in the Bayesian lensing reconstruction.
Indeed, the dipole patterns have zero mean, and should therefore not contribute significantly to Fourier scales larger than their support.
We therefore expect only a small-scale foreground bias to Bayesian lensing.

Interestingly, this intuitive picture also suggests that the bias to the lensing power spectrum should be insensitive to the fact that foregrounds and the true lensing field are correlated.
Indeed, the spurious convergence dipole patterns are oriented along the unlensed CMB gradient.
Therefore, while the spurious convergence dipoles are colocated with the true convergence monopoles, the random orientation of the spurious dipoles makes the lensing bias uncorrelated with the true lensing field.
In other words, writing the reconstructed lensing field as 
\beq
\hat{\kappa}
=
\kappa_\text{true}
+
\kappa_\text{bias}
+
\text{noise},
\eeq
we expect $\langle \kappa_\text{true} \kappa_\text{bias} \rangle = 0$,
such that the bias to the lensing power spectrum is simply 
$\langle \kappa_\text{bias}\kappa_\text{bias} \rangle$.

We quantitatively verify this intuition in Sec.~\ref{sec:results}, where we show that the lensing bias is unchanged when we destroy the correlation between true lensing and foregrounds, by shuffling the foreground cutouts but not the lensing cutouts across the patches.

\subsubsection{QE intuition}
\label{sec.intuition_qe}

\begin{figure}
\centering
\includegraphics[width=\columnwidth]{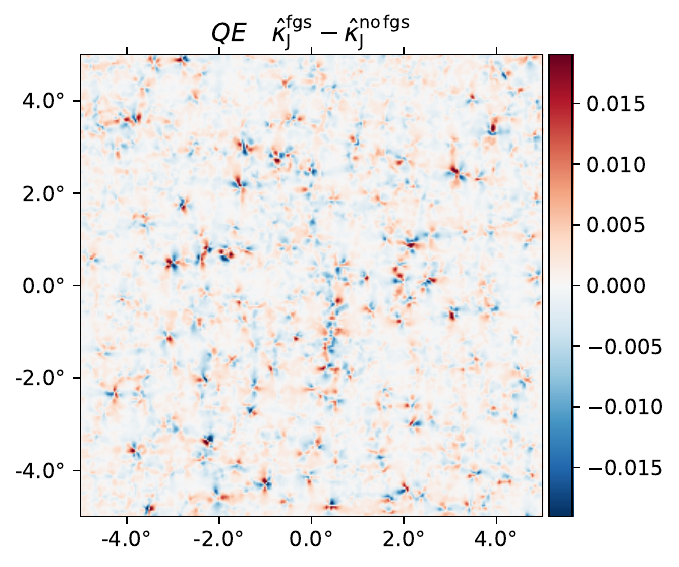}
 \caption{The difference of the reconstructed CMB lensing map in the absence of foregrounds and a reconstructed CMB lensing maps with foregrounds present for the EB estimator in QE. As explained in the main text, the EB estimator is pure shear and hence no large scale foreground bias is present; only small-scale dipole patterns appear.}
\label{fig:QE}
\end{figure}

\begin{figure}
\centering
\includegraphics[width=0.8\columnwidth]
{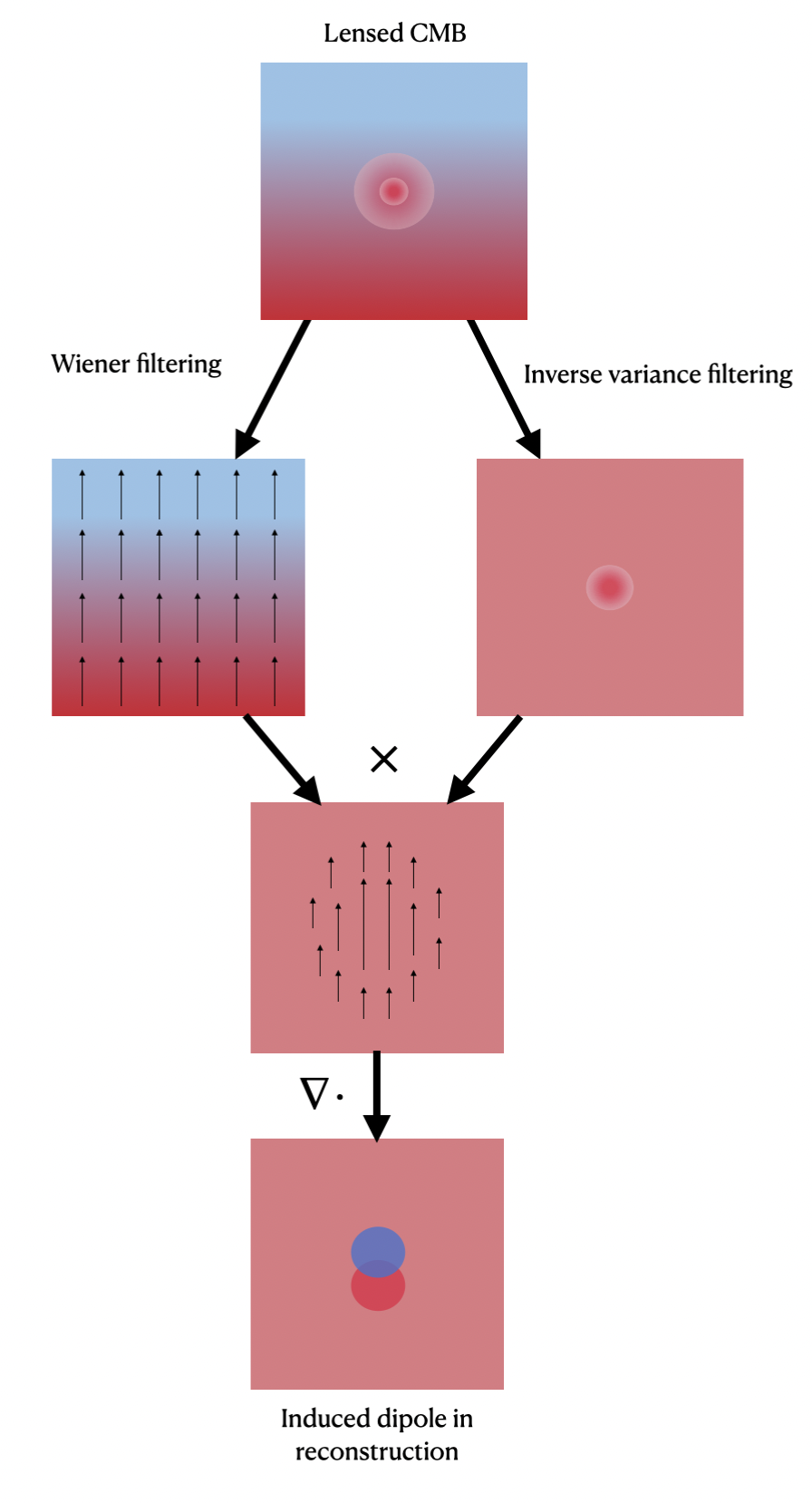}
 \caption{
 Schematic diagram of how in the formalism of the quadratic estimator on the regime of small scale lensing one also expects to obtain spurious dipoles at the location of foreground point sources in the reconstructed mass maps. Starting with a CMB map with a positive (red) point source in the center, the quadratic estimator require as input two filtered version of the CMB map: The Wiener filtered map preserves only the large scale gradient of the input CMB map, the arrows depict this point from the hot spot to the cold spot, while the inverse variance filtered map preserves the feature of the point source. The divergence of the product of the above maps generates a dipole due to the modulation of the large scale gradient by the small scale point source. 
 }
\label{fig:intuition_QE_dipoles}
\end{figure}

On small-scales, the intuition that foreground point sources produce dipoles in the reconstructed convergence field should hold for the QE as well, as visualized in Fig.~\ref{fig:intuition_QE_dipoles}.
Here, the product of the Wiener filtered large scale gradient map and the inverse variance filtered map naturally results in a dipole aligned along the gradient of the unlensed CMB, just like for Bayesian lensing.

The above intuition on small scales contrasts with the effect on large scales. 
Indeed, the QE reconstructs large-scale lensing from the observed large-scale variations in the locally-measured 2D power spectrum.
The changes to the local power spectrum can be approximated as magnification and shear, and the QE measures them by looking for a monopole and quadrupole angular dependence in the local power spectrum \cite{Schaan19,PhysRevD.108.063518,carron2024spherical}. 
Because of the very red slope of the CMB power spectrum, a positive lensing convergence produces a net reduction of power on small scales.
Foregrounds, on the other hand, add to the small-scale power. 
The QE thus mistakenly interprets this excess power as a negative lensing convergence.
As a result, the lensing bias from foregrounds is anti-correlated with the true lensing field.
This matches the result of \citep{PhysRevD.107.023504}, where the bias to the lensing power spectrum is dominated by  the correlation between true and spurious lensing and the primary and secondary bispectra dominate over the trispectrum term. 

However, in the case of the lensing QE from EB, this large-scale lensing bias is negligible, as the EB QE is effectively a shear-only estimator, i.e. its angular dependence is orthogonal to that of azimuthally-symmetric foreground point sources \cite{Sailer23}. 
This is confirmed in Fig \ref{fig:QE}, where we do not see a large-scale foreground bias, only a small-scale dipole-like bias for the reconstruction with the EB estimator in QE. In appendix \ref{app:bayqe}, we show similar reconstructions using temperature maps where the dominant bias is the  large-scale lensing bias.

\section{Results: Bias to CMB lensing from polarized sources}\label{sec:results}

We present here the main results of this paper, i.e. the size of the lensing bias in different scenarios.
As described above, the data realizations $d$ are drawn from the realistic simulations, including the non-Gaussian foreground maps (polarized point sources), whose positions are realistically correlated with the true lensing convergence field.
In contrast, the model simulations $d'$ either include no foregrounds at all, or model the foregrounds as Gaussian random fields.
Furthermore, we investigate the impact on the lensing bias \eqref{eq:bias} when the assumed foreground power spectrum in the model realizations $d'$ does not match exactly that of the realistic foregrounds simulations in $d'$.

In what follows, We often present biases in a square grid format, varying flux-cut limit and $\ell_\mathrm{max}$, along with the 1-$\sigma$ uncertainty on our estimate of the bias from bootstrapping over the 40 patches. 
We do not show the unrealistic case where no foreground mask/template subtraction would be applied, since in this extreme case it is difficult to get convergence in the reconstruction algorithm, suggesting that a large bias is incurred in the scenario of no foreground mitigation.

Fig.~\ref{fig.lensing_bias_diagram} shows the scale dependence of the bias obtained for a flux cut of $10$ mJy with an S4-deep setting for both radio and infrared sources. 
We show that the overall bias is small, even in the case where the bias would be expected to be largest, mainly 90~GHz for radio, and 148~GHz only for IR sources.
Quantitatively, a large bias is induced on small scales if the model simulations used $d^\prime$ do not include any foregrounds. 
This bias is effectively mitigated by simply including Gaussian foregrounds in the model simulations with the same power spectrum.

We estimate the corresponding biases to the overall amplitude of the CMB lensing power spectrum $\Delta{A_L}$, in units of its statistical uncertainty $\sigma$ given by

\begin{equation}
    \Delta{A^{\phi}}=\left(\sum_L\frac{{\langle \Delta \hClpp \rangle }}{\sigma^2_L}\middle/ \sum_L{\frac{1}{\sigma^2_L}}\right)\sqrt{\sum_L{\frac{1}{\sigma^2_L}}}
\end{equation}

\begin{figure*}
        \centering
        \includegraphics[width=\linewidth]{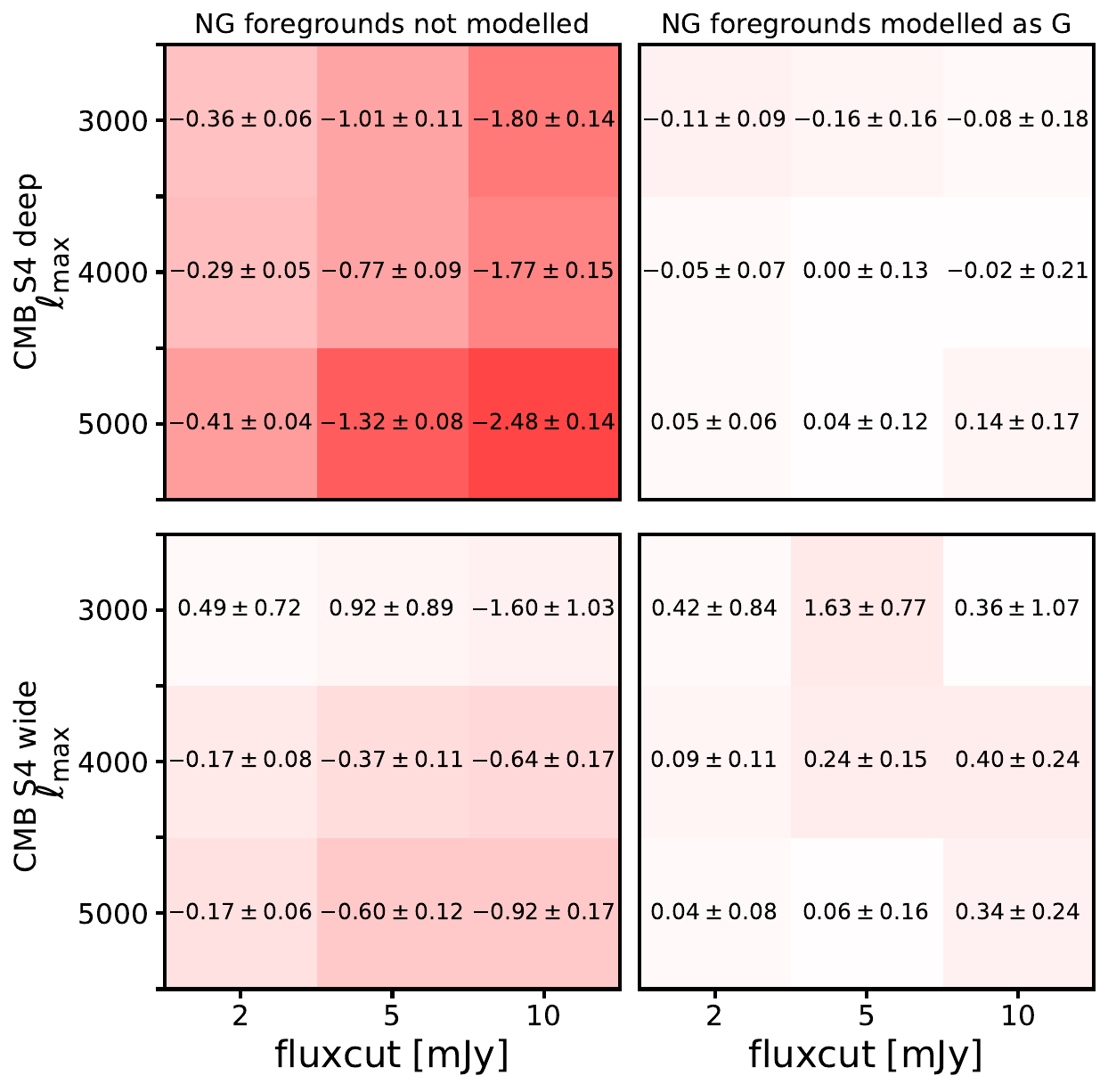}
        \caption{  Radio point source bias to the lensing power spectrum, for CMB-S4 deep (top row) and deep settings (bottom row) at $90$ GHz where radio sources are more prominent. For each setting considered, we vary the mask threshold (horizontal axis) and CMB scale cut (vertical axis). On the left we show the effect of having non Gaussian foregrounds present in the data but the foregrounds are not modelled in the simulations. This scenario, corresponding to no foreground mitigation, can incur large biases, specially on the SE-deep setting. A basic mitigation is then to include and model the simulations as Gaussian foregrounds (2nd column), this is effective in reducing the bias due to radio sources.}
     \label{fig:bias_radio}
\end{figure*}

\begin{figure*}
        \centering
        \includegraphics[width=\linewidth]{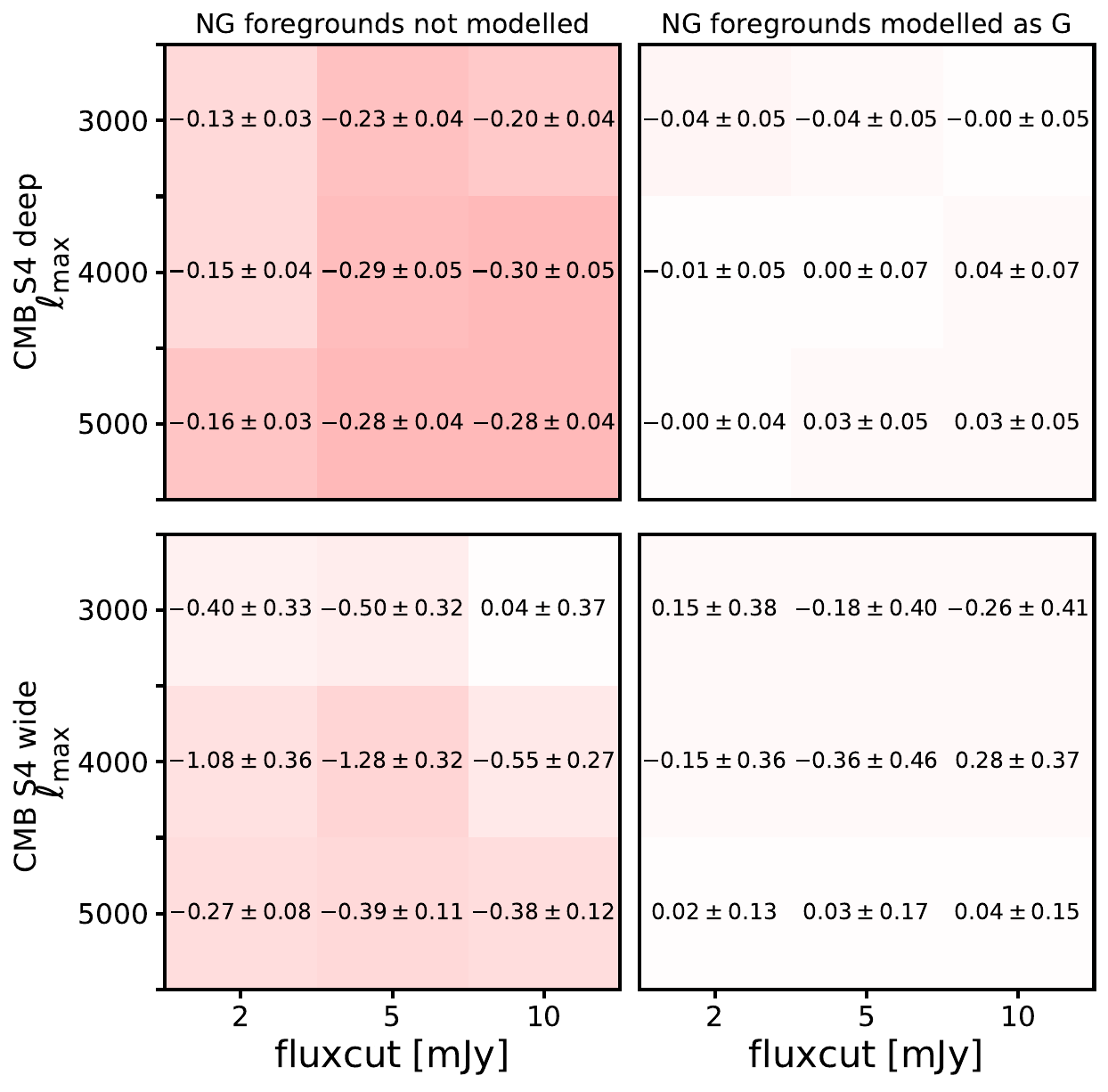}
        \caption{Infrared point-source bias to the lensing power spectrum, for CMB-S4 deep (top row) and deep settings (bottom row) at $90$ GHz where radio sources are more prominent. For each setting considered, we vary the mask threshold (horizontal axis) and CMB scale cut (vertical axis). On the left we show the effect of having non Gaussian foregrounds present in the data but the foregrounds are not modelled in the simulations. This scenario, corresponding to no foreground mitigation, can incur large biases, specially on the SE-deep setting. A basic mitigation is then to include and model the simulations as Gaussian foregrounds (2nd column), this is effective in reducing the bias due to radio sources.}
     \label{fig:bias_IR}
\end{figure*}

Fig.~\ref{fig:bias_radio} and~\ref{fig:bias_IR}
show the results for radio and IR sources, both for S4-wide and S4-deep settings, color-coded by the signal-to-noise, with larger biases shown in red.

As expected, for model simulations without any foregrounds, the lensing bias increases as we increase both the CMB maximum multipole and the flux cut used.
Restricting the scales and the masking threshold helps in reducing foreground contamination but even the most conservative treatment of $\ell_\mathrm{max}=3000$ and a flux cut of $2$ mJy still results in a $\sim-2\sigma$ bias for radio sources in S4-deep and $\sim-0.6\sigma$ bias for IR sources. 
Instead, including Gaussian foregrounds in the model simulations is effective in reducing the bias to $\leq0.3\sigma$ for most analysis choices (rightmost panels in Fig.~\ref{fig:bias_radio} and~\ref{fig:bias_IR}).

\begin{figure}
\centering
\includegraphics[width=0.9\columnwidth]
{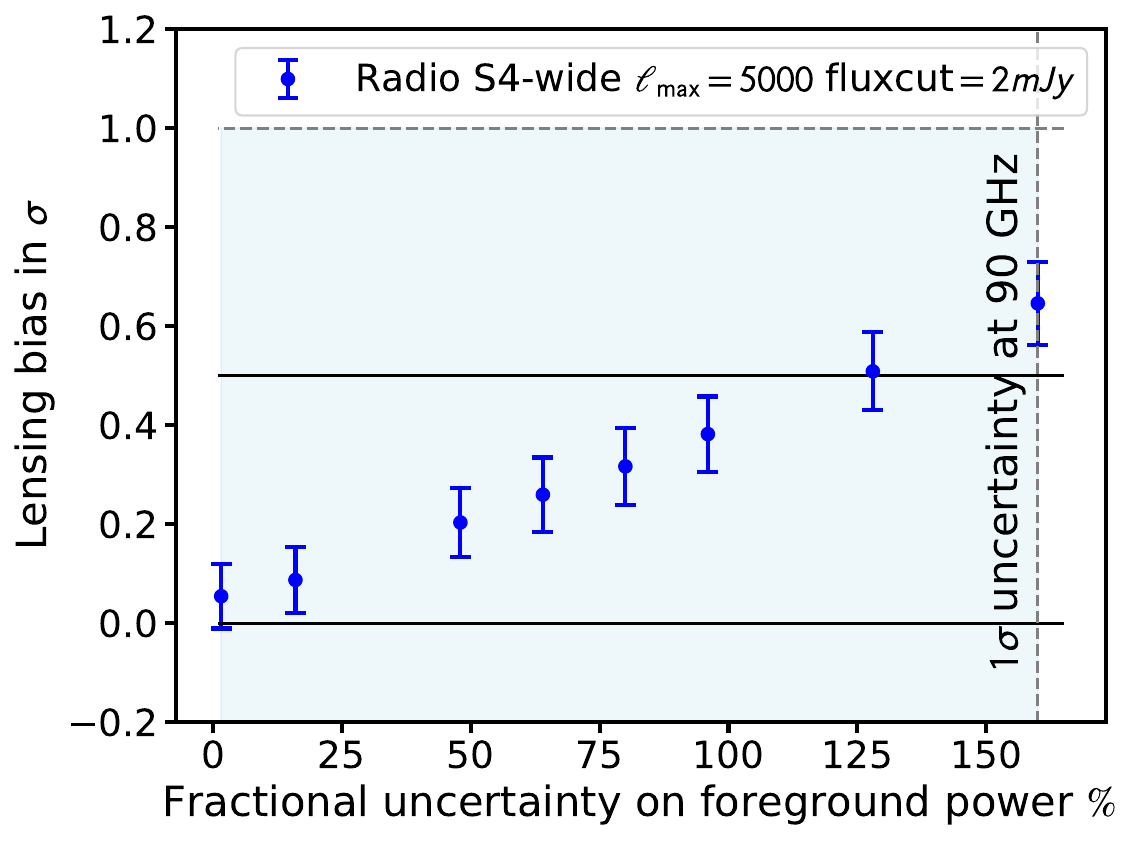}
 \caption{
 Lensing bias induced as a function of the fractional uncertainty on the radio foreground power. These biases are calculated by perturbing the standard simulations $d^\prime$ with Gaussian foregrounds with an extra amount of foreground power given by the value on the x-axis. As a guide, the $160\%$ shown on the right corresponds to the $1\sigma$ uncertainty in the foreground power measured at $90$ GHz. For CIB sources (not shown here) most of the bias is already kept below $1\sigma$ irrespective of the knowledge of the foreground power.}
\label{fig:foreground_power}
\end{figure}

In a real data analysis however, our knowledge of the true foreground power spectrum is imperfect, such that the Gaussian foreground realizations in the model may have a slightly inaccurate power spectrum.
We investigate this by rescaling the power spectrum amplitude assumed in the foreground model.
The resulting Gaussian foreground realizations in the data model thus have a slightly wrong power spectrum.
Fig. \ref{fig:foreground_power} shows the resulting lensing bias as a function of the fractional uncertainty on the foreground power spectrum.
We focus here on radio sources only, as the bias due to IR sources is too small.
We see in Fig.~\ref{fig:foreground_power} that to achieve a bias below $0.5\sigma$, one only need to control the polarized radio power uncertainty below $125\%$. 
This requirement is already achievable from the precision of the $90$GHz channel around and should be easily improved from combining multiple frequency maps of CMB-S4.
Indeed, the steep scaling of radio emission as $\propto\nu^{-3}$ will dramatically help at low frequency, despite the potentially worse resolution and noise level at these frequencies.

\begin{figure*}
        \centering
        \includegraphics[width=\linewidth]{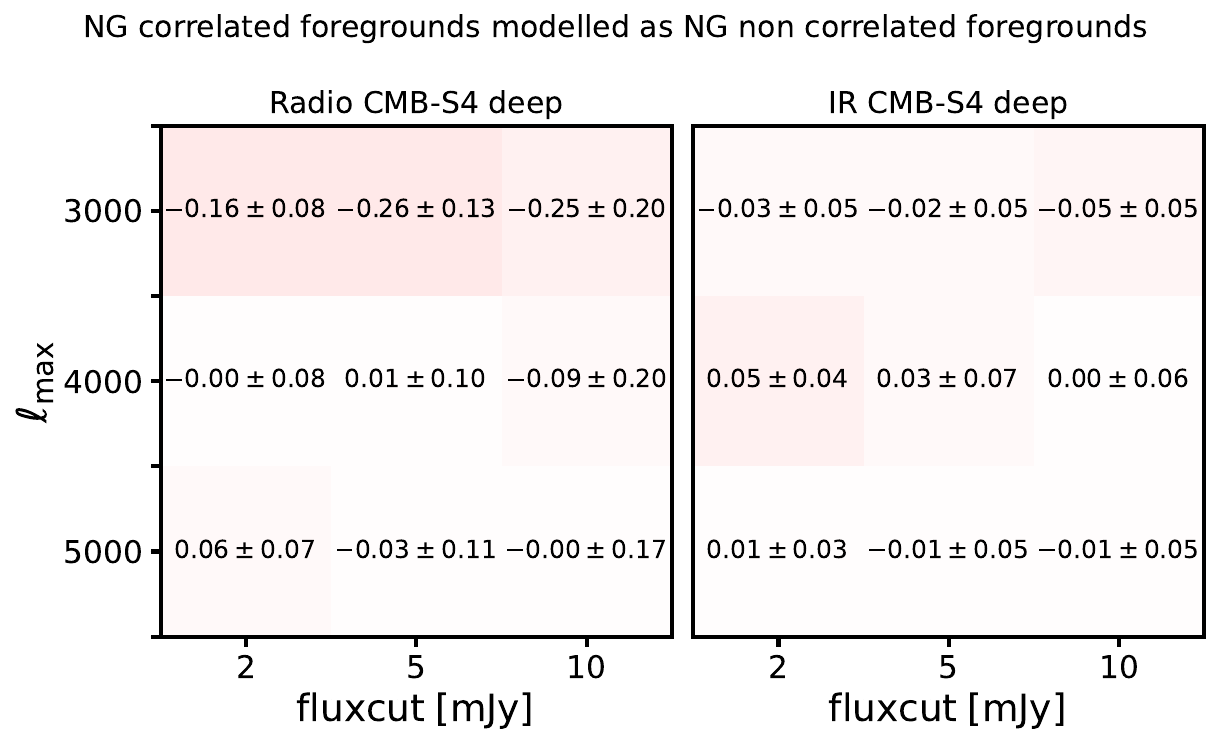}
        \caption{  Bias when the simulations $d^\prime$ used to analyse the data contain foreground realizations that are uncorrelated with the lensing field compared to the data realization that contains correlated foregrounds  as a function of $\ell_\mathrm{max}$, y-axis and fluxcut (x-axis) for radio sources, (left panel) and IR sources (right panel) for the S4-deep survey, similar results consistent with zero bias are obtained with S4-wide. This shows that the main bias contribution from foregrounds is uncorrelated with the true lensing field. }
     \label{fig:bias_radio_corr_uncorr}
\end{figure*}

Finally, we verify the intuition of Sec.~\ref{sec.psdp} that the correlation between polarized point sources and the true lensing field is irrelevant to the Bayesian lensing bias.
Fig.~\ref{fig:bias_radio_corr_uncorr} 
shows that the lensing bias is unchanged when the realistic point source maps are shuffled across cutouts, such that their correlation with the true CMB lensing convergence is nulled.

For all choices of scale cut and flux cut, the difference in  biases between correlated and uncorrelated cases are statistically consistent with zero. 
Similar results are obtained for the S4-wide settings.
This is in agreement with the intuition above, i.e. that the dipole-like bias to lensing is uncorrelated with the true lensing, since the dipoles are oriented according to the (statistically independent)primary CMB gradient.

\section{Marginalizing over a Poisson foreground component}
\label{sec:marginalize_sources}

After masking, the polarized radio and IR sources do not produce a large lensing bias, hence no further mitigation is required for CMB-S4.
For more futuristic experiments with higher sensitivity or resolution (e.g., \cite{sehgal2019cmbhd}), the foreground bias could potentially be larger.
In that case, it could be reduced further by explicitly marginalizing over a foreground model in the Bayesian analysis.
The posterior function Eq.~\eqref{eq:jointposterior} thus acquires an additional term for each foreground source $s$:
\beq
\bal
&\mathcal{P} \big(\, f, \phi, s, C_L^{\phi\phi} \, | \, d \, \big)
\propto 
\frac{\exp\left\{ -\cfrac{\left[d - \op{M} \, \op{B} \, (\Len[\phi] \, f + s) \right]^2}{2 \,\Cn} \right\}}{\det \Cn^{\nicefrac{1}{2}}} \\
&\quad\frac{\exp\left\{ -\cfrac{f^2}{2\,\Cf} \right\}}{\det  \Cf^{\nicefrac{1}{2}}} \;
\frac{\exp\left\{ -\cfrac{\phi^2}{2\,\Cphi[\Clpp]} \right\}}{\det  \Cphi[\Clpp]^{\nicefrac{1}{2}}}\;
\mathcal{P}\big(s\big),
\eal
\eeq
where $\mathcal{P}\big(s\big)$ is a fiducial prior on the statistics of the foreground field.
For a general non-Gaussian foreground field, the prior may not be known, making this approach intractable.
This is generally the case for temperature foregrounds, whose statistical properties are complex.
However, in the case of polarized radio and IR sources, the statistics is well described by a Poissonian (unclustered) set of sources, with a known flux distribution function.
In this case, the prior $\mathcal{P}\big(s\big)$ can be derived analytically (See App.~\ref{app:fg_prior}).
Given this prior, we can jointly sample the foreground field $s$, thereby marginalizing over this polarized foreground emission. 
We leave a practical implementation of this foreground marginalization to future work.

This approach is very similar in spirit to the source bias hardening estimator \cite{Osborne14, Namikawa13, plancklensing2013, Sailer20}, which modifies the lensing QE to null any response to a Poisson source field first proposed in \cite{Osborne14, Namikawa13} and applied to real data in \cite{plancklensing2013}.
In both cases, the assumption of an underlying Poissonian distribution of sources is key, as it determines the full distribution of the foreground field.
Given the effectiveness of this approach for the bias-hardened lensing QE \cite{Sailer20}, one may be optimistic that it would work well in Bayesian lensing as well.

\section{Conclusions}
\label{sec:conclusions}

This paper investigates the impact of extragalactic polarized radio and IR sources on Bayesian lensing with a CMB-S4 like setting. 
While the effect of extragalactic foregrounds on the standard QE has been studied extensively, this is the first study in the case of Bayesian lensing.
Since most of the lensing information from CMB-S4 should come from polarization, and this is also the regime where Bayesian lensing outperforms the QE the most, we have focused on this regime: polarization-only Bayesian lensing.
There, the foregrounds are particularly simple: Poisson-like sources with random polarization directions.
We therefore produce realistic simulated maps of the polarized sources, building upon the simulations from \cite{Sehgal10}.
We show how the bias to Bayesian lensing can be estimated simply, by contrasting the statistics of realistic non-Gaussian ``data simulations'' with simpler ``model simulations''.

We find that polarized IR sources do not cause a significant bias after masking the brightest individually-detected ones.
Polarized radio sources, on the other hand, produce a significant lensing bias after masking, if not accounted for.
Simply including their power spectrum in the model is sufficient to reduce the lensing bias to less than $0.5\sigma$, as long as their power spectrum is known to better than $30\%$ accuracy.
A target like this is achievable for an S4 setting where the lower frequency channels should enable the radio power spectra to be measured at or lower the required level of precision.
This result is extremely encouraging, as Bayesian lensing will outperform the QE for CMB-S4.

Despite the complexity of Bayesian lensing reconstruction, the properties of the foreground lensing bias can be understood intuitively.
Bright point sources in the Q and U maps can only be fit by the likelihood via spurious dipoles in the reconstructed convergence map, oriented along the unlensed CMB gradient.
This random orientation of the dipoles makes the lensing bias insensitive to the correlation between the point sources and the true lensing field.
This feature is also seen in the EB QE, unlike in temperature-only lensing (both Bayesian and QE).

For a futuristic CMB experiment with higher resolution or sensitivity, this foreground bias may become significant.
However, since the statistics of polarized sources is well described by a Poisson distribution, we show how such a simple non-Gaussian foreground can be included analytically in the likelihood, and thus marginalized over exactly.
Since this is not needed for CMB-S4, we leave a practical implementation of this method to future work.

We have not explored the potential bias to Bayesian lensing from Milky Way dust.
Ref.~\cite{Challinor18} (see their Table~1) studied this bias for the QE.
In polarization, they conclude that Galactic dust is an important contributor to the total map power spectrum, thus enhancing the lensing noise.
However, when including its power spectrum in the total map power in the lensing weights, no lensing bias is detected, up to the percent precision of their analysis.
A careful analysis of Milky Way dust bias to Bayesian lensing would be particularly interesting \citep{Belkner_2024}.
Bayesian lensing also outperforms the quadratic estimator in temperature, for low enough noise levels.
Quantifying the impact of temperature foregrounds on Bayesian lensing would also be useful.
We leave these explorations to future work.\\

\acknowledgments

We thank Zeeshan Ahmed, Federico Bianchini, Simone Ferraro, Mathew Madhavacheril, Sigurd Naess, Uros Seljak, Blake Sherwin, Kimmy Wu for useful discussions.
FJQ acknowledge support from the European Research Council (ERC) under the European Union’s Horizon 2020 research and innovation programme (Grant agreement No. 851274). 
FJQ is grateful for Dr Anil Seal and Trinity College Henry Barlow Scholarship for their support. 
Computations were performed on the Niagara supercomputer at the SciNet HPC Consortium. SciNet is funded by Innovation, Science and Economic Development Canada; the Digital Research Alliance of Canada; the Ontario Research Fund: Research Excellence; and the University of Toronto.
This work received support from the U.S. Department of Energy under contract number DE-AC02-76SF00515 to SLAC National Accelerator Laboratory.
This research used resources of the National Energy Research Scientific Computing Center (NERSC), a U.S. Department of Energy Office of Science User Facility located at Lawrence Berkeley National Laboratory.


\bibliographystyle{prsty}
\bibliography{refs,refs2}

\newpage
\appendix


\section{Pipeline verification} \label{app:lensing_bias}

\begin{figure}
\centering
\includegraphics[width=\columnwidth]
{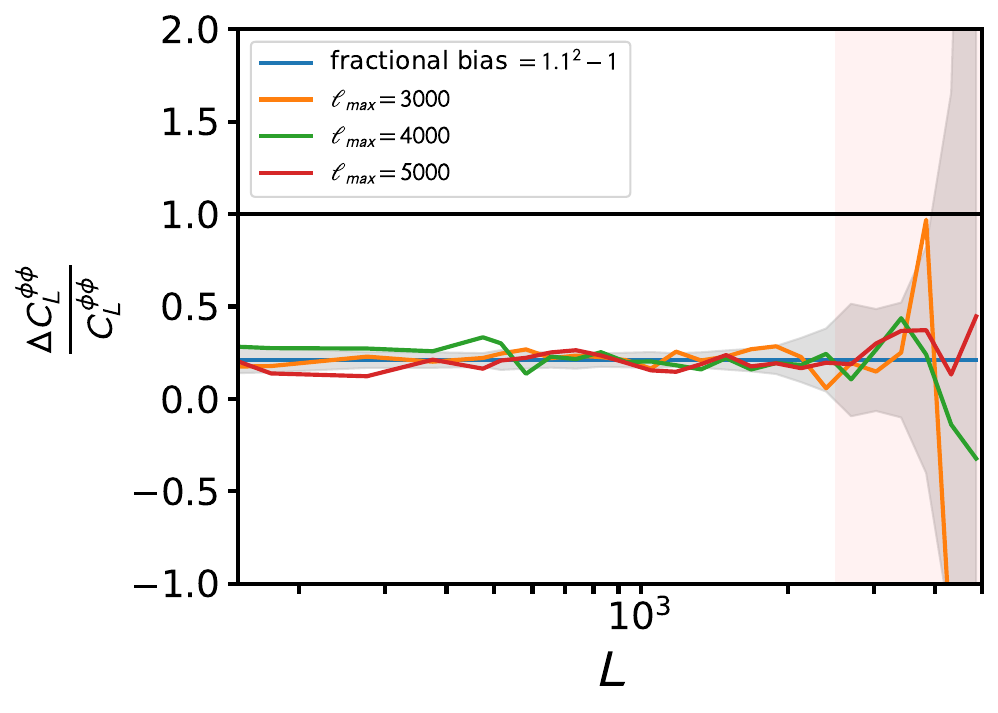}
 \caption{
 Fractional bias induced by the presence of extra lensing in the data. This artificial bias is obtained by scaling the lensing field used in each of the 400 lensed CMB simulations by $10\%$. We see that the induced bias at the four point (orange) matches well with the difference between the scaled and non-scaled lensing power spectrum in blue for all $\ell_\mathrm{max}$ used for the reconstruction.
 The red shaded region with $L>2500$ consists of the lensing scales discarded in the analyses where the SNR saturates for the CMB-S4 settings used and the bias for $\ell_\mathrm{max}=3000$ starts deviating from the expected level. The grey bands corresponds to the lensing noise per L-bin.
 }
\label{fig:LB}
\end{figure}

\begin{figure}
\centering
\includegraphics[width=\columnwidth]
{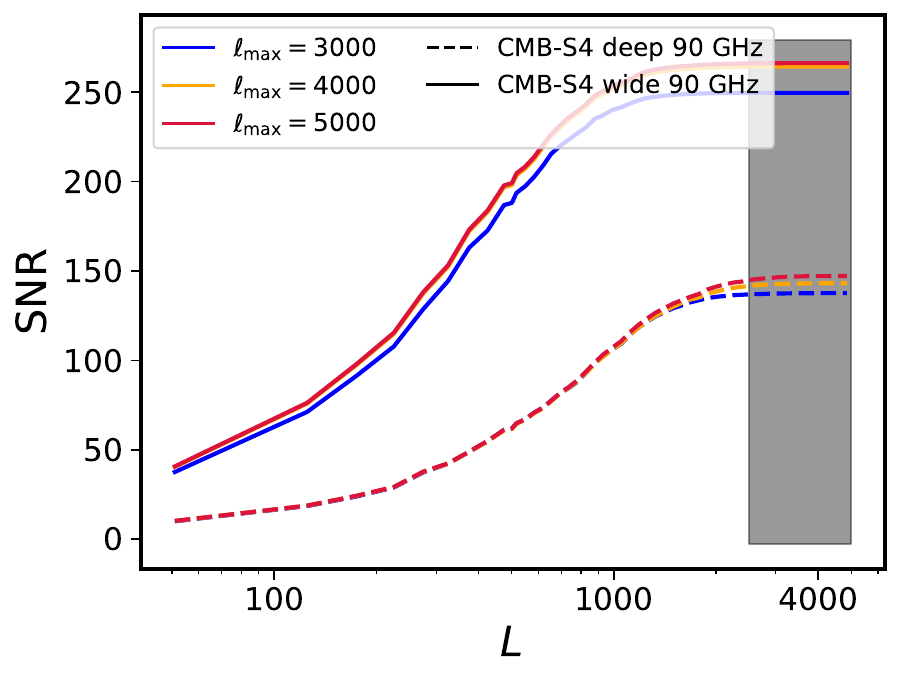}
 \caption{
Signal to noise as a function of the maximum lensing scale $\L$ used for this analysis for the CMB-S4 deep and wide settings and the $\ell_\mathrm{max}$ used. The SNR saturates at around $L=2000$ for both the deep and wide settings. Shown in grey the scales we do not consider in this work
 }
\label{fig:snr}
\end{figure}

We perform a check that the pipeline correctly measures a foreground bias when foreground contamination is present by generating 400 foreground free CMB realizations in the same manner as prescribed in section \ref{sec: simulation} but with lensing realizations scaled by $10\%$ higher compared to the usual lensing simulations. 
In Fig.  \ref{fig:LB} we show  the measured bias for a S4-deep setting with $\ell_\mathrm{max}=3000,4000,5000$ and we see that we successfully recover a bias the size of $1.1^2-1$ times the lensing power spectrum when compared to the data that do not have the lensing field scaled up for the scales used in our analysis corresponding to the white region in the plot. We use a scale cut of $L=2500$, as beyond that point the SNR saturates, as seen in Fig. \ref{fig:snr}.

\label{fig:lensing_bias}

\section{Convergence at the MAP estimates}
\label{app:convergence_steps}

\begin{figure}
\centering
\includegraphics[width=\columnwidth]
{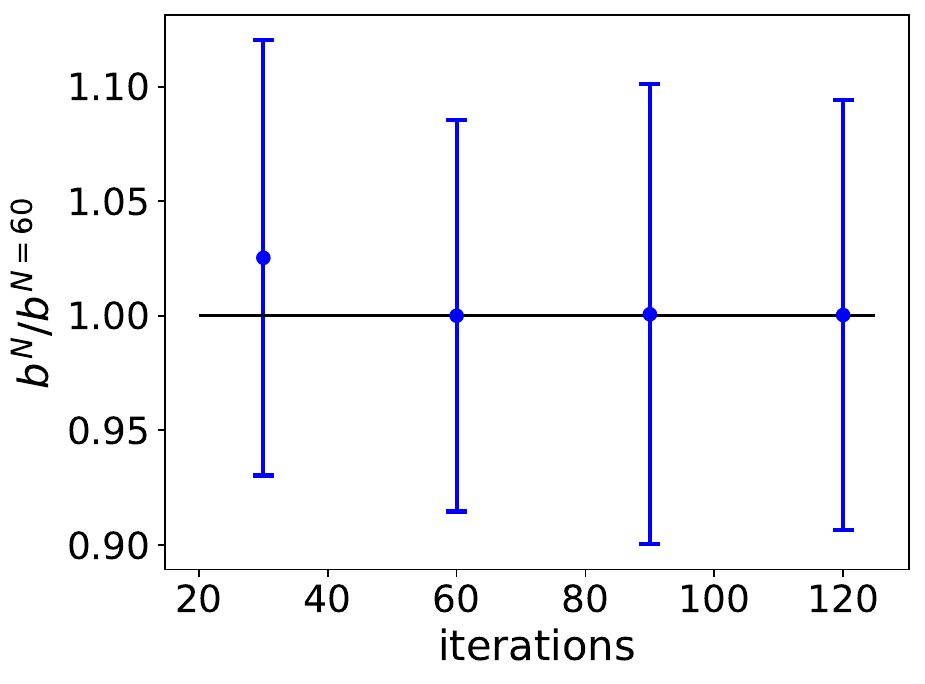}
 \caption{
 Biases  as a function of the number of iterations used normalized by the bias calculated with the baseline number of iterations $N=60$.   We adopt a conservative  $N=60$ steps for all calculations in the reported results, even though with $N=30$ steps, the results are already converged.
}
\label{fig:CB}
\end{figure}

We verify that the reported results are converged with $N=60$, (the number of iteration steps used by the maximiser to estimate the maximum a posteriori), by computing the bias for radio sources with $\ell_\mathrm{max}=5000$ and fluxcut$=10$ mJ using a higher iteration setting with $N=90$ and $N=120$ . In Fig. \ref{fig:CB} we see that even for $N=30$ steps, the value we obtain is consistent with the baseline result of using $N=60$ steps, showing that the results with $N=60$ are well converged.

\section{Foreground bias intuition from reconstruction using temperature maps}\label{app:bayqe}

\begin{figure*}
  \centering
  \begin{minipage}{.5\textwidth}
    \centering
    \includegraphics[width=.9\linewidth]{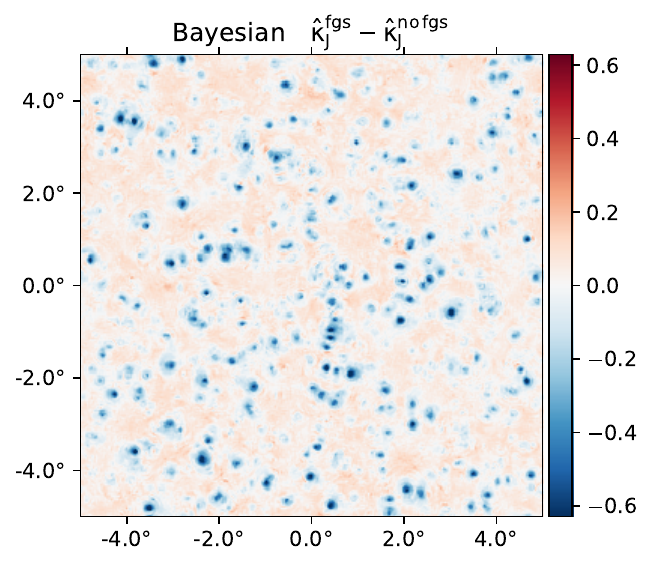}
  \end{minipage}%
  \begin{minipage}{.5\textwidth}
    \centering
    \includegraphics[width=.9\linewidth]{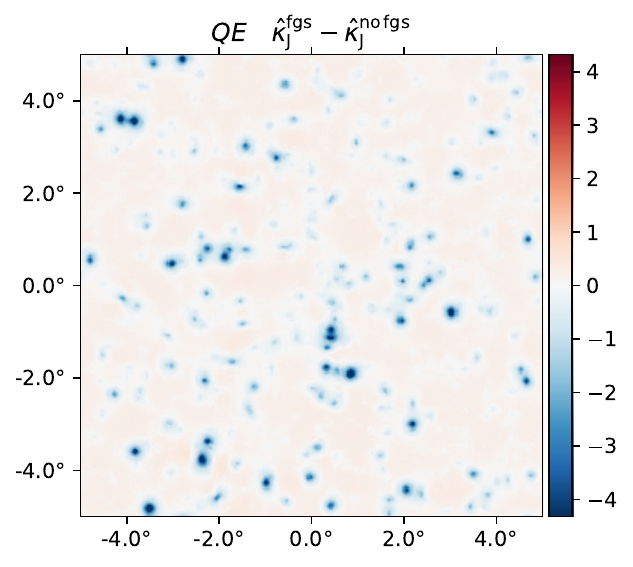}
  \end{minipage}
    \caption{Lensing bias for bayesian (left) and QE (right) for reconstruction done on CMB temperature maps.}
\label{fig.temperature_qe}
\end{figure*}

We show that when using CMB temperature maps to reconstruct lensing instead of polarization, the dominant contribution to the bias comes from the effect of foreground on large lensing scales. 
Namely, we see a negative correlation on the position of the point sources instead of the small-lensing scale dipole pattern discussed in the main text for polarization. 
This holds true for both Bayesian lensing and QE as can be seen in Fig. \ref{fig.temperature_qe}.
This can be understood from the effect of foregrounds on the local (isotropic) 2D power spectrum.
Indeed, excess foregrounds enhance the local temperature power spectrum.
On the other hand, a positive large-scale convergence rescales the local power spectrum (magnification effect).
Because the CMB is redder than $\propto \ell^{-2}$, this rescaling is a reduction in the local power spectrum.
Thus, both QE and Bayesian lensing incorrectly assign the excess foreground power to a negative convergence ~\cite{Schaan19,PhysRevD.108.063518}.

\section{Map prior for polarized Poissonian sources}
\label{app:fg_prior}

In this section, we derive the map PDF for Poisson sources, such as the radio and IR galaxies.
For simplicity, we start by deriving the PDF of the source temperature field $\mathcal{P}\big( T\big)$, then extend the derivation to the PDF of the source polarization $\mathcal{P}\big( Q, U\big)$.

In what follows, we assume that the number of sources per pixel within a given flux $S$ bin is Poissonian, with a mean determined by the source luminosity function $dN/ dS d\Omega$, where $d\Omega$ is a unit sold angle.
For the polarized maps, we shall assume that the polarized fraction $\alpha$ is uniform across all sources, and that the polarization angles are uniformly distributed, independently across sources.

\subsection{Temperature}

To derive the PDF $\mathcal{P}\big( T\big)$, we think of the temperature map as a specific surface brightness map
$T\sim I$ (with units [power/area/solid angle/frequency]).
The intensity observed in a given pixel receives a contribution from each source within the pixel:
\beq
I = \int dS\ \frac{dN}{dSd\Omega}\ S.
\eeq
We rewrite as the sum
$I = \int dI$, 
where
$dI = \left( S/\Omega_p \right) dN$
is the intensity contributed by sources with flux $S \pm dS$.
Here $\Omega_p$ is the pixel solid angle,
and $dN$ is the number of sources with flux $S \pm dS$ found within the pixel.
It is a Poisson variable with mean determined by the source flux distribution:
$d\bar{N} = dS \Omega_p \left( dN / dSd\Omega\right)$.
In other words:
\beq
\mathcal{P}\left( dN \right)
=
\frac{d\bar{N}^N}{N!} e^{-d\bar{N}}.
\eeq
Because we are considering sums of random variables, 
it is advantageous to work in terms of the characteristic functions $\tilde{\mathcal{P}}(\tilde{x})$, i.e. the Fourier transforms of the probability distribution functions $\mathcal{P}(x)$.
For the Poisson variable $dN$, it is:
\beq
\tilde{\mathcal{P}} \left( \tilde{dN} \right)
=
\exp \left[
d\bar{N}
\left(
e^{i \tilde{dN}} - 1
\right)
\right]
.
\eeq
Since $dI$ and $dN$ are linearly related as $dI = (S/\Omega_p) dN$,
the characteristic function of $dI$ is simply related to that of $dN$ as
\beq
\tilde{\mathcal{P}}_{\tilde{dI}} \left( \tilde{dI} \right)
=
\tilde{\mathcal{P}}_{\tilde{dN}} \left( \tilde{dN} = \tilde{dI} (S/\Omega_p) \right)
,
\eeq
hence:
\beq
\tilde{\mathcal{P}} \left( \tilde{dI} \right)
=
\exp \left[
d\bar{N}
\left(
e^{i (S/\Omega_p)\tilde{dI}} - 1
\right)
\right]
.
\eeq
Since the total intensity is the sum of the elemental intensities $dI$, 
the PDF of $I$ is the convolution of the PDFs of the $dI$.
In Fourier space, this convolution of PDFs simply becomes a product of characteristic functions:
\beq
\mathcal{P}_I = \circledast \mathcal{P}_{dI}
\quad\rightarrow\quad
\tilde{\mathcal{P}}_{\tilde{I}} \left(\tilde{I}\right) = \prod \tilde{\mathcal{P}}_{\tilde{dI}} \left(\tilde{I}\right),
\eeq
Hence:
\beq
\bal
\tilde{\mathcal{P}} \left(\tilde{I}\right)
&=
\exp \left[
\int d\bar{N}\ 
\left(
e^{i (S/\Omega_p)\tilde{I}} - 1
\right)
\right]\\
&=
\exp \left[
\Omega_p
\int dS \frac{dN}{dS d\Omega}\ 
\left(
e^{i (S/\Omega_p)\tilde{I}} - 1
\right)
\right].\\
\eal
\eeq
Inverse Fourier-transforming the characteristic function gives us the desired PDF:
\beq
\bal
\mathcal{P}\left( I \right)
&=
\int_0^\infty \frac{d\tilde{I}}{\left( 2\pi \right)}\
\exp \left[
iI\tilde{I}
+
\right.\\
&\left.
\Omega_p
\int_{S_\text{flux cut}}^\infty dS \frac{dN}{dS d\Omega}\ 
\left(
e^{i (S/\Omega_p)\tilde{I}} - 1
\right)
\right]
.\\
\eal
\eeq
This formula is indeed dimensionally correct: 
$\tilde{I}$ is the inverse of an intensity, whereas $I$ and $S/\Omega_p$ are intensities.
Furthermore, the PDF depends explicitly on the pixel size $\Omega_p$.
Indeed, as the pixel size increases, the number of sources in the pixel increases, such as the fractional Poisson fluctuations decrease.
At the limit $\Omega_p \rightarrow \infty$, we indeed get
$\mathcal{P}\left( I \right)
\rightarrow 
\delta^D\left(  
I - \int_{S_\text{flux cut}}^\infty dS \frac{dN}{dS d\Omega}\ S
\right).$

\subsection{Q \& U polarizations}

Having derived the temperature PDF for Poissonian sources, we now turn to the joint PDF of the Stokes $Q$ and $U$.
Let $P = \sqrt{Q^2 + U^2}$ be the polarized intensity and $\theta$ the polarization angle, such that
\beq
\left\{
\bal
& Q = P\ \text{cos}\ 2\theta\\
& U = P\ \text{sin}\ 2\theta\\
\eal
\right.
.
\eeq

\paragraph{Single source with fixed polarized amplitude}

For a given source, we assume the polarized intensity and angle to be statistically independent,
i.e.
$\mathcal{P}(P, \theta) = \mathcal{P}(P)\ \mathcal{P}(\theta)$.
We assume the polarization angle to be uniformly distributed in $[0,\pi]$, and we start by considering a set of sources with fixed polarized intensity $P=P_0$.
Hence
\beq
\mathcal{P}(P, \theta)
=
\delta^D\left( P-P_0 \right)\
\frac{\mathbb{I}_{\theta \in [0,\pi]}}{\pi}.
\eeq
As above, we work in terms of characteristic functions, better suited to treating sums of random variables.
\beq
\bal
\tilde{\mathcal{P}} \left( \tilde{Q}, \tilde{U} \right)
&=
\int dQ dU\
\mathcal{P} \left( Q, U \right)
e^{-i Q\tilde{Q}}
e^{-i U\tilde{U}}\\
&=
\int dP d\theta\
\mathcal{P} \left( P, \theta \right)
e^{-i Q\tilde{Q}}
e^{-i U\tilde{U}}\\
&=
\int dP\
\delta^D\left( P-P_0 \right)\times\\&
\int d\theta\
\frac{\mathbb{I}_{\theta \in [0,\pi]}}{\pi}
e^{-i \tilde{Q} P \text{cos}2\theta}
e^{-i \tilde{U} P \text{sin}2\theta}\\
&=
\int_0^\pi \frac{d\theta}{\pi}\
e^{-iP\left[ \tilde{Q} \text{cos}2\theta + \tilde{U} \text{sin}2\theta \right]}\\
&=
\int_0^\pi \frac{d\theta}{\pi}\
e^{-iP\tilde{P}\text{cos}\left( 2\theta + \phi_{(\tilde{Q}, \tilde{U})} \right)}\\
\eal
\eeq
hence the PDF is given by the Bessel function of the first kind:
\beq
\tilde{\mathcal{P}} \left( \tilde{Q}, \tilde{U} \right)
=
J_0\left( P_0 \tilde{P} \right)
\quad\text{with}\quad
\tilde{P} \equiv \sqrt{\tilde{Q}^2 + \tilde{U}^2}.
\eeq

\paragraph{$N$ sources with fixed polarized intensity}

The Stokes parameters are additive for incoherent radiation.
Thus for $N$ sources, $Q=Q_1+...+Q_N$ and $U=U_1+...+U_N$, and the PDF of $(Q, U)$
is the convolution of all the individual PDFs.
The characteristic function is thus simply:
\beq
\tilde{\mathcal{P}}_N \left( \tilde{Q}, \tilde{U} \right)
=
J_0\left( P_0 \tilde{P} \right)^N
\quad\text{with}\quad
\tilde{P} \equiv \sqrt{\tilde{Q}^2 + \tilde{U}^2}.
\eeq

This result also holds for zero sources ($N=0$), since in that case $\tilde{\mathcal{P}}_0 \left( \tilde{Q}, \tilde{U} \right)=1$,
i.e.
$\mathcal{P}_0 \left( Q, U \right) = \delta^D(Q) \delta^D(U)$.

\paragraph{Poisson number of sources with fixed polarized intensity}

If the number $N$ of sources within the pixel is Poisson distributed with mean $\bar{N}$, the PDF of $(Q, U)$ becomes:
\beq
\mathcal{P} \left( Q, U \right)
=
\sum_{N\geq 0}
\mathcal{P}_N \left( Q, U \right)
\text{Poisson}_{\bar{N}}(N).
\eeq
In terms of characteristic functions, this becomes
\beq
\bal
\tilde{\mathcal{P}} \left( \tilde{Q}, \tilde{U} \right)
&=
\sum_{N\geq 0}
\tilde{\mathcal{P}}_N \left( \tilde{Q}, \tilde{U} \right)
\text{Poisson}_{\bar{N}}(N)\\
&=
\sum_{N\geq 0}
J_0\left( P_0 \tilde{P} \right)^N
\frac{\bar{N}^N}{N!} e^{-\bar{N}}\\
&=
e^{-\bar{N}}
\sum_{N\geq 0}
\frac{\left[\bar{N} J_0\left( P_0 \tilde{P} \right)\right]^N}{N!}\\
&=
e^{-\bar{N}}
e^{\bar{N} J_0\left( P_0 \tilde{P} \right)}
\eal
\eeq
i.e.:
\beq
\tilde{\mathcal{P}} \left( \tilde{Q}, \tilde{U} \right)
=
e^{\bar{N}\left[ J_0\left( P_0 \tilde{P} \right) -1 \right]}.
\eeq

\paragraph{Population of sources with variable polarized intensity}

We can now tackle the more general case, where the sources have different polarized intensities.
We assume that the polarization fraction $\alpha$ is constant for all sources, such that the polarized intensity is 
$P = \alpha I = \alpha S / \Omega_p$, 
where $\Omega_p$ is again the pixel angular size.

The number $dN$ of sources with flux $S\pm dS$, i.e with polarized intensity $P \pm dP$, within a pixel, is assumed to be Poissonian, with mean 
$d\bar{N} = dS \Omega_p dN / dSd\Omega$.
These sources contribute $(dQ, dU)$ to the pixel, with characteristic function:
\beq
\tilde{\mathcal{P}}\left( \tilde{dQ}, \tilde{dU} \right)
=
e^{d\bar{N}\left[ J_0\left( P \tilde{dP} \right) -1 \right]}.
\eeq

Adding together all the $(dQ, dU)$ from all the $S$ or $P$ bins corresponding to multiplying the corresponding characteristic functions:
\beq
\bal
\tilde{\mathcal{P}}_{\left( \tilde{Q}, \tilde{U} \right)}
&=
\prod
\tilde{\mathcal{P}}_{\left( \tilde{dQ}, \tilde{dU} \right)}\\
&=
\prod
e^{d\bar{N}\left[ J_0\left( P \tilde{P} \right) -1 \right]}\\
&=
e^{\int d\bar{N}\left[ J_0\left( P \tilde{P} \right) -1 \right]}\\
\eal
\eeq
i.e.:
\beq
\tilde{\mathcal{P}}_{\left( \tilde{Q}, \tilde{U} \right)}
=
e^{\Omega_p\int dS \frac{dN}{dS d\Omega} \left[ J_0\left( \alpha (S/\Omega_p) \tilde{P} \right) -1 \right]}.
\eeq
Fourier transforming finally gives the PDF:
\beq
\bal
\tilde{\mathcal{P}}_{\left( Q, U \right)}
&=
\int_0^\infty \frac{\tilde{P} \tilde{dP}}{2\pi}\ 
J_0(P\tilde{P})\\
&e^{\Omega_p\int_{S_\text{flux cut}}^\infty dS \frac{dN}{dS d\Omega} \left[ J_0\left( \alpha (S/\Omega_p) \tilde{P} \right) -1 \right]}.\\
\eal
\eeq
As in the temperature case, the dimensional correctness of this equation is easily verified.
Furthermore, as the pixel size increases, and more and more sources average out inside the pixel, 
the PDF reduces to the expected results:
$\mathcal{P}\left( Q, U \right) \rightarrow \delta^D(Q) \delta^D(U)$,
i.e. 
$\tilde{\mathcal{P}}\left( \tilde{Q}, \tilde{U} \right) \rightarrow 1$.

\end{document}